\documentclass[pdflatex,sn-mathphys-num]{sn-jnl}

\usepackage{graphicx}%
\usepackage{multirow}%
\usepackage{amsmath,amssymb,amsfonts}%
\usepackage{amsthm}%
\usepackage{mathrsfs}%
\usepackage[title]{appendix}%
\usepackage{xcolor}%
\usepackage{textcomp}%
\usepackage{manyfoot}%
\usepackage{booktabs}%
\usepackage{algorithm}%
\usepackage{algorithmicx}%
\usepackage{algpseudocode}%
\usepackage{listings}%
\usepackage{tikz}
\usetikzlibrary{shapes.geometric, arrows}
\usepackage{array}
\usepackage{longtable}
\usepackage{tabularx}
\usepackage{lscape}

\tikzstyle{startstop} = [rectangle, rounded corners, minimum width=3cm, minimum height=1cm,text centered, draw=black, fill=red!30]
\tikzstyle{process} = [rectangle, minimum width=3cm, minimum height=1cm, text centered, draw=black, fill=orange!30]
\tikzstyle{decision} = [diamond, minimum width=3cm, minimum height=1cm, text centered, draw=black, fill=green!30]
\tikzstyle{arrow} = [thick,->,>=stealth]

\begin{document}

\title[Enhancing Class Diagram Dynamics: A Natural Language Approach with ChatGPT]{Enhancing Class Diagram Dynamics: \\A Natural Language Approach with ChatGPT}

\author*{\fnm{ROUABHIA} \sur{Djaber}}\email{djaber.rouabhia@univ-tebessa.dz}

\author{\fnm{HADJADJ} \sur{Ismail}}\email{ismail.hadjadj@univ-tebessa.dz}

\affil{\orgdiv{Computer Science Department}, \orgname{Chahid Cheikh Laarbi Tebessi University}, \city{Tebessa}, \postcode{12000}, \state{Algeria}}

\abstract{

Integrating artificial intelligence (AI) into software engineering can transform traditional practices by enhancing efficiency, accuracy, and innovation. This study explores using ChatGPT, an advanced AI language model, to enhance UML class diagrams dynamically, an underexplored area. Traditionally, creating and maintaining class diagrams are manual, time-consuming, and error-prone processes. This research leverages natural language processing (NLP) techniques to automate the extraction of methods and interactions from detailed use case tables and integrate them into class diagrams.

The methodology involves several steps: (1) developing detailed natural language use case tables by master's degree students for a "Waste Recycling Platform," (2) creating an initial static class diagram based on these tables, (3) iteratively enriching the class diagram through ChatGPT integration to analyze use cases and suggest methods, (4) reviewing and incorporating these methods into the class diagram, and (5) dynamically updating the PlantUML \cite{plantuml} class diagram, followed by evaluation and refinement. Findings indicate that the AI-driven approach significantly improves the accuracy and completeness of the class diagram. Additionally, dynamic enhancement aligns well with Agile development practices, facilitating rapid iterations and continuous improvement.

Key contributions include demonstrating the feasibility and benefits of integrating AI into software modeling tasks, providing a comprehensive representation of system behaviors and interactions, and highlighting AI's potential to streamline and improve existing software engineering processes. Future research should address identified limitations and explore AI applications in other software models.

}


\maketitle
\newpage
\section{Introduction}
Recent advancements in artificial intelligence (AI) have profoundly impacted various domains, including software engineering \cite{Ali2023,batarseh2021}. AI technologies are now employed in multiple areas of software development, such as requirements engineering, automated testing, and defect prediction \cite{Zhao2021,batarseh2021,Wang2023}. The integration of AI, particularly natural language processing (NLP), is anticipated to revolutionize traditional software engineering practices by enhancing efficiency, accuracy, and innovation \cite{Martinez-Fernandez2022, Khurana2023}.

Unified modeling language (UML) class diagrams are essential in software engineering for designing and documenting systems, they offer a structured representation of a system’s components, their attributes, methods, and relationships \cite{Berardi2005, Gray2020}. Traditionally, creating and maintaining UML class diagrams is a manual, labor-intensive process, often resulting in time-consuming and error-prone tasks, leading to inconsistencies and inaccuracies throughout the software development lifecycle \cite{Sajjii2023}. However, advancements in AI, particularly NLP, provide opportunities to automate and improve this process.

AI-driven approaches have been applied to various aspects of software development, including requirements engineering, code generation, and system modeling \cite{Wang2023}. One critical area where AI has shown significant potential is in creating UML class diagrams \cite{Sajjii2023}. Capturing dynamic aspects, such as interactions and behaviors within a system, is crucial for a comprehensive understanding of system functionality and for creating accurate and maintainable designs. Including methods in class diagrams provides detailed specifications of class functionalities, enhancing documentation, design clarity, and maintainability while supporting modularity and object-oriented design principles.

Despite these advancements, existing automated tools often struggle with the intricate details and dynamic aspects of complex systems \cite{Ahmed2022}. A critical gap remains in the literature: the dynamic enhancement of UML class diagrams using AI to capture interactions and behaviors within a system. Current methodologies predominantly focus on static components, overlooking dynamic aspects.

Translating use case scenarios into UML class diagrams manually is error-prone and inconsistent \cite{Anda2005}. Developers must interpret use case scenarios, identify relevant methods, and integrate these methods into class diagrams. This process is time-consuming and susceptible to human error, leading to incomplete or inaccurate diagrams. Additionally, maintaining these diagrams to reflect changes in requirements or system functionalities adds complexity and effort. While intermediary models such as sequence diagrams, state diagrams, or activity diagrams provide detailed understanding, their creation requires additional expertise and time, complicating the overall development process.

Integrating AI into the dynamic enhancement of UML class diagrams through NLP represents a significant advancement in the field. This study explores using ChatGPT, an advanced AI language model developed by OpenAI \cite{OpenAI2021}, to enhance the accuracy and completeness of UML class diagrams. ChatGPT's natural language understanding and generation capabilities make it a valuable tool for analyzing use case scenarios and iteratively incorporating extracted insights into class diagrams. By leveraging ChatGPT, we aim to automate the identification and integration of methods into class diagrams, transforming them from static representations into dynamic models that accurately reflect both the structure and behavior of the system, providing a more comprehensive system representation.

The objectives of this study are threefold:
\begin{enumerate}
\item \textbf{Automate Method Identification:} Utilize ChatGPT to analyze natural language use case tables and identify relevant methods for each class.
\item \textbf{Dynamic Diagram Enhancement:} Iteratively update the UML class diagrams to include identified methods, ensuring the diagrams evolve to accurately represent the complete functionality of the system.
\item \textbf{Evaluate Effectiveness:} Assess the accuracy, efficiency, and comprehensiveness of the AI-enhanced class diagrams compared to the initial ones.
\end{enumerate}

To guide this study, we formulated the following research questions:
\begin{itemize}
\item \textbf{RQ1:} To what extent is it feasible to integrate ChatGPT to dynamically enhance the class diagram dynamics effectively?
\item \textbf{RQ2:} How does the AI-driven enhancement of UML class diagrams impact the efficiency and accuracy of software modeling compared to traditional methods?
\item \textbf{RQ3:} What are the limitations and challenges of using ChatGPT for dynamic class diagram enhancement, and how can these be mitigated?
\end{itemize}

This study focuses on an exemplary use case, a "Waste Recycling Platform," developed by master's degree students in computer science. The PlantUML code representing the initial class diagram for this platform includes core classes, attributes, and basic relationships but lacks detailed methods. The study leverages 23 detailed use case tables formulated in natural language, structured with columns such as Use Case ID, Actor, Description, Pre-condition, Trigger, Main Scenario, Post-condition, and Exceptions.

The findings indicate that the AI-driven approach significantly improves the accuracy and completeness of class diagrams while reducing the manual effort required. The dynamic enhancement aligns well with Agile development practices, facilitating rapid iterations and continuous improvement. This methodology not only provides a more comprehensive representation of system behaviors and interactions but also streamlines the process of maintaining and updating class diagrams.

While the findings demonstrate significant benefits, the methodology's effectiveness depends on the quality of the input use case tables and may require human oversight to ensure accuracy. Additionally, the study uses a single use case for validation, which may affect the generalizability of the results. Future research will focus on addressing these limitations and expanding the methodology’s applicability to broader and more complex systems.

Integrating AI, particularly NLP, into the software modeling process represents a significant advancement in software engineering practices. By automating routine tasks such as method extraction and incorporation, AI can significantly reduce manual effort, minimize human error, and ensure more accurate and comprehensive class diagrams. This study not only demonstrates the feasibility and benefits of such an integration but also paves the way for further research and development in AI-assisted software engineering methodologies.

In summary, this study presents a novel approach to dynamically enhancing UML class diagrams by leveraging ChatGPT's capabilities. The findings highlight the significant improvements in the accuracy, efficiency, and comprehensiveness of class diagrams, demonstrating the transformative potential of AI in software engineering.

The remainder of this paper is structured as follows: Section 2 details the methodology used in this study, including the initial class diagram, use case tables, analysis, and dynamic incorporation. Sections 3 and 4 present the results and discuss the implications of our findings. Section 5 concludes the paper and suggests directions for future research.

\section{Literature Review}

The integration of artificial intelligence (AI) into software engineering has garnered significant attention, with numerous studies highlighting its potential to transform traditional practices \cite{Martinez-Fernandez2022}. One critical area where AI has shown promise is in creating UML class diagrams, which are pivotal for visualizing system structures and dynamics \cite{Berardi2005, Gray2020}. Historically, the creation and maintenance of class diagrams have been manual processes, often resulting in time-consuming, error-prone, and resource-intensive tasks \cite{Berardi2005}. These traditional methods struggle to keep pace with the complexities of modern software systems, leading to inconsistencies and inaccuracies that can cascade through the software development lifecycle \cite{Sajjii2023}.

\subsection{AI and NLP in Software Engineering}

To address these challenges, various automated and semi-automated tools have been developed. The advent of AI, particularly natural language processing (NLP), has introduced new paradigms in software engineering. AI-driven approaches have been applied to numerous aspects of software development, including requirements engineering, code generation, and system modeling \cite{Wang2023}. For instance, Zhao et al. \cite{Zhao2021} highlighted the potential of NLP in automating requirements engineering processes, showcasing significant efficiency and accuracy gains.

Pre-trained language models, such as those developed by OpenAI, have demonstrated significant capabilities in understanding and generating human-like text, which can be leveraged to automate code generation tasks and improve the consistency of generated artifacts \cite{Perez2021}. This presents a transformative opportunity for bridging the gap between natural language specifications and software artifacts.

\subsection{Methods to Enhance Class Diagrams}

Several studies have proposed methods to enhance class diagrams. Elena \cite{Elena2014} suggests analyzing behavioral models represented by collaboration diagrams to define relationships between class diagram constituents. Similarly, Egyed \cite{Egyed2002} introduces automated abstraction techniques to simplify complex structures, while Guéhéneuc \cite{Gueheneuc2004} discusses reverse engineering tools for building precise class diagrams from Java programs. Huang et al. \cite{Huang2016} suggest using knowledge graphs to provide additional context and facilitate better abstraction.

Eichelberger \cite{Eichelberger2002} presents aesthetic criteria and layout algorithms to standardize and enhance the visual quality of UML class diagrams. Anda and Sjøberg \cite{Anda2005} investigate the role of use cases in constructing class diagrams and how different techniques affect the quality of the resulting diagrams.  Sharma et al. \cite{Sharma2010} introduce a dependency analysis-based approach to derive UML class diagrams automatically from natural language requirements, addressing challenges of manual intervention and pre-processing.

Osman et al. \cite{Osman2014} study machine learning techniques to condense class diagrams, highlighting key classes. Similarly, Alashqar \cite{Alashqar2021} presents an approach for the automatic generation of UML diagrams from scenario-based user requirements, showcasing the efficiency of NLP in processing user requirements. Nasiri et al. \cite{Nasiri2020} propose methods for generating class diagrams from user stories in agile methodologies, emphasizing the integration of dynamic and static aspects to enhance accuracy. Kumar and Sanyal \cite{Kumar2008} develop the SUGAR tool for generating static UML models from requirement analysis, simplifying the translation of requirements into visual models. Lucassen et al. \cite{Lucassen2017} explore the extraction of conceptual models from user stories, leveraging visual narrators to improve the accuracy and relevance of generated UML class diagrams.

\subsection{AI and Optimization Techniques}

Several researchers have explored AI and optimization techniques to improve UML class diagrams. Tantithamthavorn et al. \cite{Tantithamthavorn2020} emphasize the importance of explainable AI in software engineering, particularly in improving software defect prediction models. Sergievskiy and Kirpichnikova \cite{Sergievskiy2018} discuss optimizing UML class diagrams using design patterns and anti-patterns, proposing a plugin for analyzing XMI files that contain class diagram descriptions. Eiglsperger et al. \cite{Eiglsperger2004} propose using the topology–shape–metrics paradigm for the automatic layout of UML class diagrams in orthogonal style, significantly improving readability and aesthetics by minimizing edge crossings and bends.

\subsection{Enhancing Consistency and Comprehension}

Ha and Kim \cite{Ha2013} propose cross-checking rules to enhance the consistency of UML diagrams, focusing on the consistency between UML static and dynamic diagrams. Their methodology emphasizes the importance of maintaining consistency without requiring complex intermediate representations.  Briand et al. \cite{Briand2012} enhance interaction testing by combining UML sequence and state machine diagrams. Aoumeur and Saake \cite{Aoumeur2002} propose integrating UML structural and behavioral diagrams using rewriting logic, enhancing consistency, and supporting rapid prototyping.

\subsection{NLP and Heuristic Rules}

Abdelnabi et al. \cite{Abdelnabi2020} explore generating UML class diagrams using NLP techniques combined with heuristic rules, addressing the complexity and computational cost issues associated with traditional approaches. Their research demonstrates how automated processes can significantly reduce human intervention and errors in diagram creation. Maatuk and Abdelnabi \cite{Maatuk2021} further the research by developing methods to generate UML use case and activity diagrams using similar NLP techniques and heuristic rules, providing a comprehensive framework for automatic diagram generation. This work emphasizes the integration of different types of UML diagrams for a cohesive modeling process. P., D. et al. \cite{P2020} enhance the extraction of business vocabularies and rules from UML use case diagrams using natural language processing, demonstrating the advanced capabilities of NLP in handling complex software engineering tasks. Their work highlights how NLP can be used to improve the precision and comprehensiveness of business models.  Their research highlights the importance of visual tools in enhancing the understanding and clarity of conceptual models.

\subsection{Consistency and Traceability}

Omer et al. \cite{Omer2021} and Dawood and Sahraoui \cite{Dawood2018} focus on the consistency and traceability of requirements and design, utilizing bi-directional traceability and natural language processing to ensure the alignment of requirements with design artifacts. Their studies emphasize the importance of maintaining consistency throughout the software development lifecycle. Arellano et al. \cite{Arellano2015} provide a framework for processing textual requirements using natural language processing, facilitating the extraction and organization of requirements into structured formats. This framework aids in managing complex requirements and translating them into formal models. Yue et al. \cite{Yue2015} introduce ATOUCAN, an automated framework that derives UML analysis models from use case models, underscoring the potential of automation in improving the consistency and completeness of UML diagrams. This framework highlights the benefits of automating analysis model generation from detailed use case descriptions.

\subsection{Reviews and Surveys}

Several comprehensive reviews and surveys have been conducted to consolidate the findings and methodologies in this domain. Salehi and Burgueño \cite{Salehi2018} review emerging AI methods in structural engineering, highlighting AI-based solutions' efficiency and accuracy over traditional approaches. Abdelnabi et al. \cite{Abdelnabi2021} provide a thorough survey of approaches for generating UML class diagrams from natural language requirements, highlighting ongoing challenges and the need for more advanced tools. Ahmed et al. \cite{Ahmed2022} conducted a systematic literature review focusing on the automatic transformation of natural language to UML. Their study emphasizes the limitations of existing approaches, including constraints on ambiguity, length, structure, anaphora, incompleteness, and atomicity of input text. They argue that while heuristic rule-based and machine learning-based approaches have shown promise, they still face significant challenges in fully automating the generation of UML diagrams. Ahmed et al. also highlight the necessity for a common dataset and evaluation framework to advance research consistently. Their findings suggest that a combination of heuristic rules and machine learning techniques could provide a more robust solution for translating natural language into UML diagrams.

\subsection{Addressing Gaps in Literature}

Despite advancements in AI applications in software engineering, there remains a gap in the literature regarding the use of AI, particularly NLP, to enhance the dynamics of class diagrams. Existing methodologies predominantly focus on static aspects, such as automating class diagram generation or leveraging AI for requirements engineering and code generation. However, the dynamic aspects of class diagrams, which capture interactions and behaviors within a system, remain underexplored and are often overlooked.

Our research attempts to address this gap by proposing a novel methodology that utilizes ChatGPT to analyze natural language use cases and integrate the extracted insights into UML class diagrams. This approach automates the extraction of methods and interactions from textual descriptions and iteratively refines the class diagrams to improve their accuracy and completeness.

\section{Methodology}

This study leverages an iterative approach to dynamically enrich a PlantUML class diagram using ChatGPT and pre-generated use case tables. The use case in focus is a "Waste Recycling Platform," with the initial natural language use case tables and class diagram developed by master's degree students in computer science.

Our objective is to automate method identification and incorporation, addressing the gap in the initial static diagram. Adding methods to class diagrams is crucial. It specifies the functionalities of each class, representing both the system's structure and behavior. This enhances the clarity, maintainability, and comprehensiveness of the diagrams.

The steps involved in our methodology are as follows:

\begin{enumerate}
    \item \textbf{Use Case Development:} Master's degree students created detailed natural language use case tables for the "Waste Recycling Platform." These tables describe various interactions and functionalities required by the system.

    \item \textbf{Initial Class Diagram:} Based on the use case tables and other artifacts, students created an initial static class diagram. This diagram includes classes, attributes, and basic relationships but lacks detailed methods.

    \item \textbf{Iterative Enrichment:}
    \begin{itemize}
        \item \textit{ChatGPT Integration:} We used ChatGPT to analyze the pre-generated use case tables and suggest relevant methods for each class. ChatGPT's natural language processing capabilities enabled the extraction of functional requirements from the use cases.
        \item \textit{Method Identification:} Suggested methods were reviewed and iteratively incorporated into the class diagram. This step involved verifying the relevance and correctness of the methods in the context of the system's functionalities.
        \item \textit{Diagram Update:} The PlantUML class diagram was dynamically updated to include the identified methods, ensuring that the diagram evolves to reflect the complete functionality of the system.
    \end{itemize}
    
    \item \textbf{Evaluation and Refinement:} The enriched class diagram was evaluated for completeness and accuracy. 
\end{enumerate}

This approach ensures that the final class diagram is not only structurally sound but also behaviorally comprehensive, providing a robust blueprint for system implementation.

\subsection{Use of PlantUML}

PlantUML was chosen for this study due to its simplicity, flexibility, and integration capabilities. As a tool that uses a simple textual description to generate diagrams, PlantUML allows for easy updates and modifications, making it ideal for iterative processes. Its support for various UML diagrams, including class diagrams, sequence diagrams, and state diagrams, provides comprehensive modeling capabilities. Furthermore, PlantUML integrates seamlessly with version control systems, allowing for efficient tracking of changes. The ability to embed PlantUML in documentation and its compatibility with various IDEs and text editors further enhance its usability, making it a powerful tool for dynamic and collaborative software design.

\subsection{Use Case Tables (Guiding Method Identification)}

This study utilizes 23 detailed use case tables (Appendix A.1 to A.23), formulated in natural language, and structured with columns like Use Case ID, Actor, Description, Pre-condition, Trigger, Main scenario, Post-condition, and Exceptions. Originally derived by master's degree students in computer science from a waste recycling platform specification document, these tables serve as the primary reference for identifying methods necessary for each class to facilitate described functionalities.

A representative subset of nine User actor use cases is provided below in Table \ref{tab}. For this example; in all scenarios, the actor is the "User," who must be logged in to perform the described actions (except for UC1).

\begin{table}[h]
\centering
\begin{tabular}{|c|l|l|l|l|}
\hline
\textbf{ID} & \textbf{Description} & \textbf{Trigger (User)} & \textbf{Main Scenario (User)} & \textbf{Post-condition} \\
\hline UC1 & Registration & accesses registration page & provides information & User account created \\
UC2 & Buying Recyclable Waste & selects a product & completes purchase & Transaction processed \\
UC3 & Selling Recyclable Waste & accesses selling page & enters sale details & Sale submitted \\
UC4 & Requesting Waste Collection & fills collection form & submits request & Collection scheduled \\
UC5 & Viewing Collection Points & accesses map/list & browses points & User views details \\
UC6 & Listing Recyclable Waste & accesses listing page & enters product details & Listing published \\
UC7 & Requesting Waste Transport & fills transport form & submits request & Transport scheduled \\
UC8 & Submitting Feedback & fills feedback form & submits feedback & Feedback processed \\
UC9 & Managing Profile & accesses profile settings & updates profile & Profile updated \\
\hline
\end{tabular}
\caption{Representative Use Cases for User Actor}
\label{tab}
\end{table}

This selection highlights the iterative integration of user-centric functionalities into the dynamically enriched class diagram, demonstrating the method’s feasibility and effectiveness in capturing requirements and translating them into actionable design elements.

\textit{\textbf{Note:}} For global clarity and comprehensive understanding, the detailed use case tables for all actors, including the representative subset provided here, are available in Appendix \ref{secA}.

\subsection{Initial Class Diagram}

The original PlantUML class diagram, developed by master's degree students in computer science, lacked methods and served solely as a static representation of the system’s structure. Despite existing class relationships, the absence of detailed methods significantly limited understanding of system functionality. By dynamically enriching the diagram with methods from detailed use case tables, we transformed it into a dynamic representation, facilitating a more complete understanding of system dynamic behaviors. This enriched diagram now serves as a valuable tool for communication, design, and implementation. Figure \ref{fig:initial-diagram} depicts the initial diagram. For clarity and to support the main narrative of the article, the initial diagram PlantUML code is included in Appendix \ref{secB}.

The primary classes include:
\begin{itemize}
    \item \textbf{Users}: Categorized as Individual and Corporate, with attributes like name, address, and registration date.
    \item \textbf{Products}: Represented by diverse materials like Plastic, Paper, and Metal, with details like quantity, price, and listing date.
    \item \textbf{Transactions}: Capturing purchase activities with buyer/seller information, quantity, and total price.
    \item \textbf{Reviews}: Enabling users to rate and comment on products.
    \item \textbf{Service Requests}: Associated with users and categorized by type and status.
    \item \textbf{Collection/Recycling Points}: Differentiated by types accepted and operating hours.
    \item \textbf{Additional entities}: Including Transport Companies, State Administration, Payment Gateways, Information Resources, and Reward Systems.
\end{itemize}

\subsection{Analysis, Identification, and Dynamic Incorporation}

ChatGPT employs various NLP techniques to discern relevant classes and actions within the use case tables. Leveraging its natural language comprehension capabilities, it analyzes each scenario to identify necessary methods for each class, ensuring a thorough grasp of system behaviors. These insights form the foundation for method definition and subsequent dynamic updates to the PlantUML code, by incorporating the identified methods.

The process maintains consistency and accuracy through version control and validation against both the use cases and the original class diagram. The outlined algorithm (Algorithm \ref{alg:dynamic-enhancement})  delineates this methodology.

\begin{algorithm}
\caption{Dynamic Class Diagram Enhancement}
\label{alg:dynamic-enhancement}
\begin{algorithmic}[1]
\State \textbf{Step 1: Use Case Analysis and Method Identification}
    
    - ChatGPT-assisted Analysis: Extract actions, behaviors, inputs, and outputs from each use case table for involved classes.
    
    - Method Definition: Define required methods for each class based on extracted information, adhering to PlantUML syntax and UML best practices.
\State \textbf{Step 2: Dynamic Diagram Update}
    
    - Generate PlantUML code and update the diagram with identified methods, preserving existing elements.
\State \textbf{Step 3: Validation and Iteration}
    
    - Validate the updated diagram against the use case and original diagram. 
    
    - Address discrepancies and iterate with new use case tables.
    
\While{New use case tables are available}

    - Repeat from Step 1.

\EndWhile
\end{algorithmic}
\end{algorithm}

\subsection{The Guiding Main Prompt}

A central "main prompt," presented below, directs the approach with specific instructions:

\begin{quote}
    \textbf{Prompt 1:} "You are assigned the task of enriching the dynamics of a given PlantUML class diagram based on detailed use cases presented in table format. Your role involves deeply analyzing each use case table to determine the interactions and behaviors necessary within the system. Utilizing the provided class diagram, identify the involved classes and define the methods or operations required for each class to support the described functionality. Generate PlantUML code to update the class diagram, ensuring adherence to syntax and UML best practices. Maintain consistency in property names and utilize only existing attributes in the class diagram for method definitions. As new use case tables are provided, dynamically update the class diagram to accurately represent the system’s behavior. Validate the completeness and accuracy of the updated diagram after each iteration. Your process must follow guidelines for thorough analysis, method incorporation, and validation, ensuring consistency throughout. Additionally, preserve all added methods in each iteration and maintain consistency in naming conventions, formatting, and UML notation."
\end{quote}

This prompt stresses thorough analysis, accurate method incorporation, and consistent validation. Following this framework ensures systematic enhancement of the class diagram, guaranteeing completeness and accuracy in representing the system’s behavior.

\subsection{Conclusion}
In summary, our methodology leverages ChatGPT to dynamically enhance UML class diagrams by analyzing natural language use cases and integrating the extracted methods and interactions. This approach automates the traditionally manual and error-prone process of class diagram creation and maintenance, improving both accuracy and efficiency. While our findings demonstrate significant benefits, the methodology's reliance on the quality of input use cases and the use of a single use case for validation highlight areas for further research and improvement. Future work will focus on addressing these limitations, expanding the methodology's applicability, and exploring the long-term impacts of AI integration in software engineering practices.

\section{Results}

This section presents the findings from the analysis and enhancement of the class diagram using ChatGPT for natural language processing of use cases. These findings highlight the methodology's effectiveness in refining system design and enhancing functional requirements. Initially, the system's class diagram contained core classes with basic relationships. After applying the ChatGPT methodology, we observed significant improvements in added methods, class structures, and modified relationships. This section provides a comprehensive overview of the enhanced system design.

\subsection{Overview of Initial and Enhanced Class Diagrams}

The initial class diagram consisted of core classes and their relationships, establishing the basic framework for entities including:
\begin{itemize}
    \item Users
    \item Products
    \item Transactions
    \item Reviews
    \item Collection points
    \item Recycling points
    \item Service requests
    \item Transport companies
    \item State administration
    \item Payment gateways
    \item Environmental impacts
    \item Information resources
    \item Reward systems
    \item Payment details
    \item Shipping details
\end{itemize}
Each class was defined with pertinent attributes, and associations were depicted to represent the interactions among these entities.

\subsubsection{Enhanced Class Diagram}

Upon implementing the methodology with ChatGPT to analyze the natural language use cases, the enhanced class diagram was developed. This diagram features additional methods and refined relationships. Methods were specifically incorporated into the User, Transaction, Review, ServiceRequest, PaymentGateway, and PlatformManager classes to meet the functional requirements derived from the use cases.

\subsection{Visualizing the Transformations}

\begin{figure}[h]
    \centering
    \includegraphics[width=\textwidth]{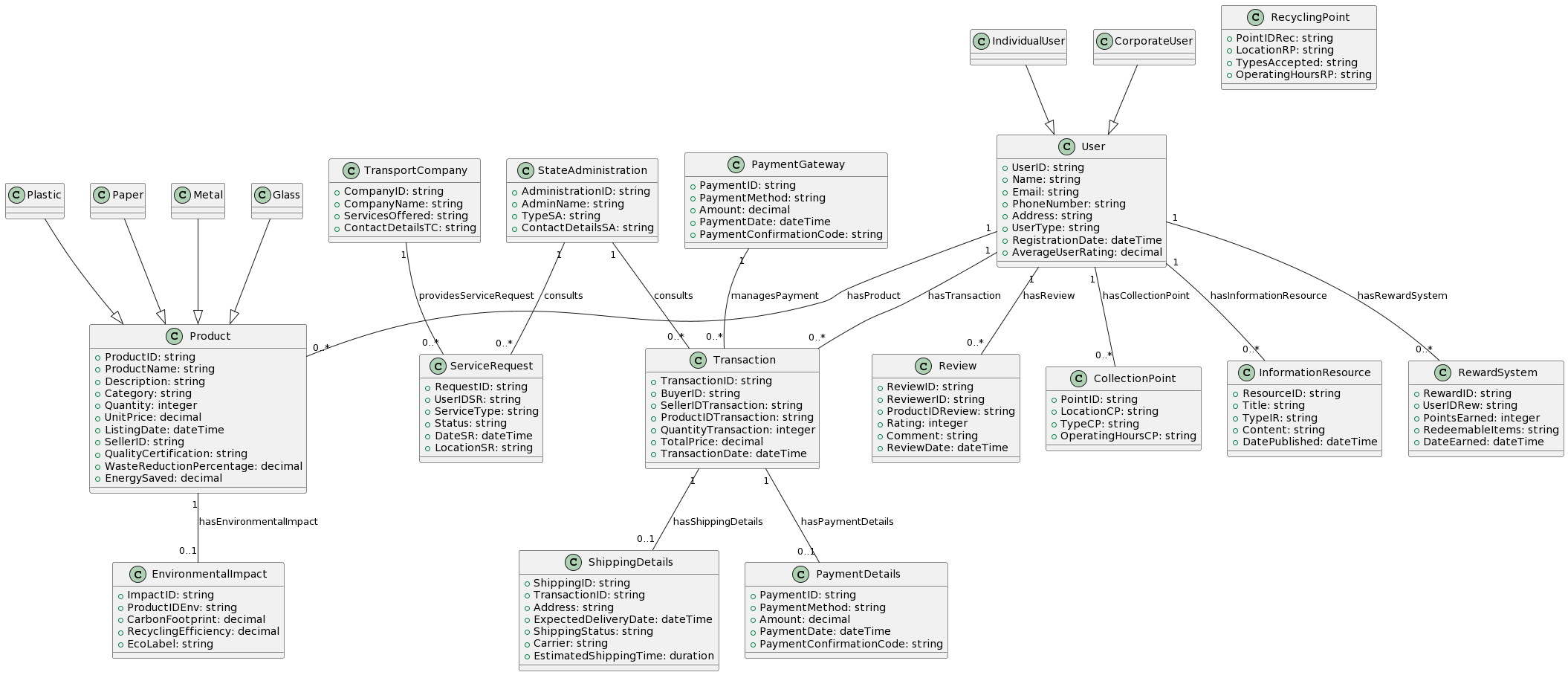}
    \caption{Initial Class Diagram}
    \label{fig:initial-diagram}
\end{figure}

\begin{figure}[h]
    \centering
    \includegraphics[width=\textwidth]{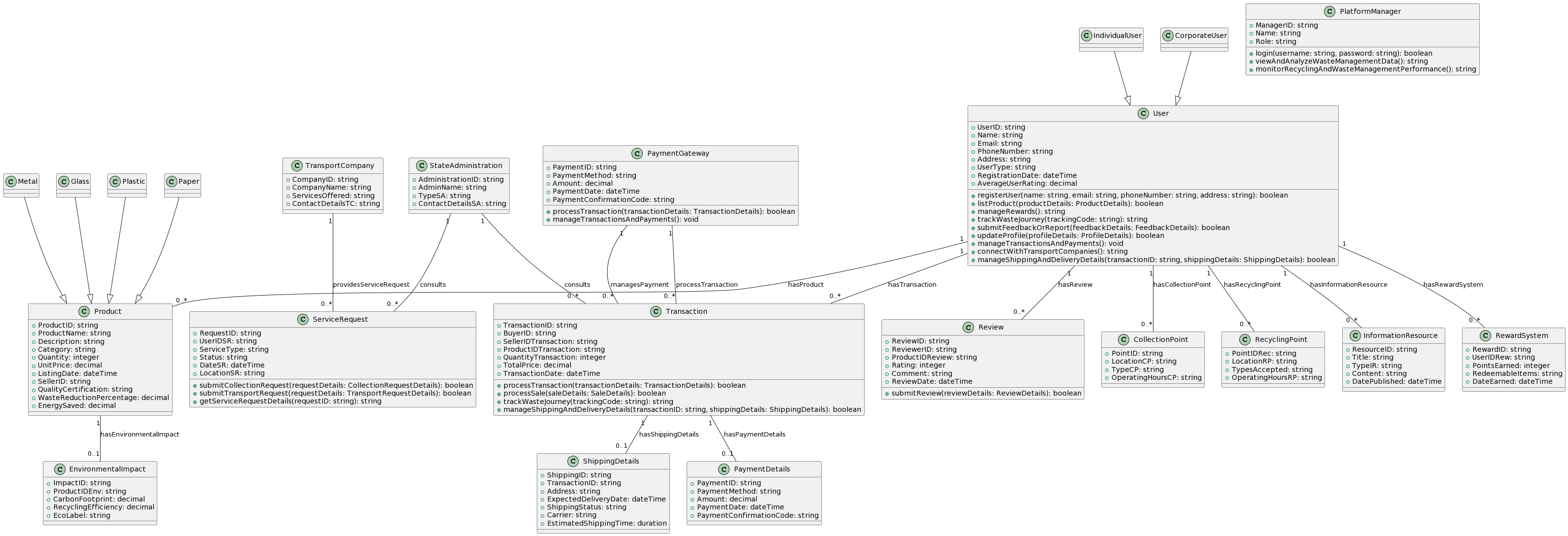}
    \caption{Enhanced Class Diagram}
    \label{fig:enhanced-diagram}
\end{figure}

The transition from the initial static class diagram to the dynamically enriched diagram illustrated substantial improvements in representing system behaviors. Figures \ref{fig:initial-diagram} and \ref{fig:enhanced-diagram} depict these transformations, highlighting the inclusion of methods and interactions that were previously missing. The enhanced diagram provides a more detailed view of the system's dynamics, facilitating improved understanding and communication among stakeholders.

\textbf{Note}: The corresponding PlantUML codes for the initial and enhanced class diagrams are provided in Appendix B and Appendix C for clarity and to maintain the main flow of the article.

\subsection{Comparative Analysis}

The enhanced class diagram provided a more functional and detailed representation of the system compared to the initial diagrams. This improvement is quantified in Table \ref{tab:comparative-analysis}, which compares the number of classes, methods, and relationships before and after the enhancement.

\begin{table}[h]
    \centering
    \begin{tabular}{|l|c|c|}
        \hline
        \textbf{Feature} & \textbf{Initial Diagram} & \textbf{Enhanced Diagram} \\
        \hline
        Number of Classes & 21 & 22 \\
        Number of Methods & 0 & 22 \\
        Number of Relationships & 19 & 21 \\
        Dynamic Behaviors Captured & No & Yes \\
        \hline
    \end{tabular}
    \caption{Comparative Analysis of Initial and Enhanced Class Diagrams}
    \label{tab:comparative-analysis}
\end{table}

\subsection{Behavioral Alignment}

The methods incorporated into the enhanced class diagram were validated against 23 distinct use cases to ensure their applicability and effectiveness. This validation confirms the precision and relevance of the AI-driven enhancements, demonstrating the role of AI in refining software design and enhancing system functionality. Each method, from user registration to managing transport services, was meticulously aligned with corresponding use case requirements, underscoring the comprehensive approach taken in this study.

\subsubsection{Added Methods to Classes}

The enhancement process introduced new methods to several key classes, ensuring that the functional requirements derived from the use cases are adequately addressed.

\paragraph{User Class}

\textbf{Key Changes and Impact:} New methods were added to enhance user interaction and management of products, rewards, waste tracking, feedback, profile updates, transactions, and shipping details.

\begin{itemize}
    \item \texttt{registerUser(name: string, email: string, phoneNumber: string, address: string): boolean} - Registers a new user with necessary details and returns a boolean indicating registration success.
    \item \texttt{listProduct(productDetails: ProductDetails): boolean} - Lists a product with provided details and returns a boolean indicating listing success.
    \item \texttt{manageRewards(): string} - Manages user rewards and returns relevant reward information.
    \item \texttt{trackWasteJourney(trackingCode: string): string} - Tracks the waste journey and returns the current status.
    \item \texttt{submitFeedbackOrReport(feedbackDetails: FeedbackDetails): boolean} - Processes user feedback or reports and returns a boolean indicating success.
    \item \texttt{updateProfile(profileDetails: ProfileDetails): boolean} - Updates user profile details and returns a boolean indicating success.
    \item \texttt{manageTransactionsAndPayments(): void} - Manages all transactions and payments for the user.
    \item \texttt{connectWithTransportCompanies(): string} - Connects the user with transport companies and returns relevant information.
    \item \texttt{manageShippingAndDeliveryDetails(transactionID: string, shippingDetails: ShippingDetails): boolean} - Manages shipping details for a transaction and returns a boolean indicating success.
\end{itemize}

\paragraph{Transaction Class}

\textbf{Key Changes and Impact:} Methods added to handle transaction processing, sale processing, waste tracking, and shipping details management.

\begin{itemize}
    \item \texttt{processTransaction(transactionDetails: TransactionDetails): boolean} - Processes a transaction and returns a boolean indicating success.
    \item \texttt{processSale(saleDetails: SaleDetails): boolean} - Processes a sale and returns a boolean indicating success.
    \item \texttt{trackWasteJourney(trackingCode: string): string} - Tracks the waste journey and returns the current status.
    \item \texttt{manageShippingAndDeliveryDetails(transactionID: string, shippingDetails: ShippingDetails): boolean} - Manages shipping details for a transaction and returns a boolean indicating success.
\end{itemize}

\paragraph{Review Class}

\textbf{Key Changes and Impact:} Method added to handle review submissions.

\begin{itemize}
    \item \texttt{submitReview(reviewDetails: ReviewDetails): boolean} - Accepts review details and returns a boolean indicating success.
\end{itemize}

\paragraph{ServiceRequest Class}

\textbf{Key Changes and Impact:} Methods added to handle collection and transport requests, and to provide request details.

\begin{itemize}
    \item \texttt{submitCollectionRequest(requestDetails: CollectionRequestDetails): boolean} - Processes a collection request and returns a boolean indicating success.
    \item \texttt{submitTransportRequest(requestDetails: TransportRequestDetails): boolean} - Processes a transport request and returns a boolean indicating success.
    \item \texttt{getServiceRequestDetails(requestID: string): string} - Provides details for a service request based on request ID.
\end{itemize}

\paragraph{PlatformManager Class}

\textbf{Key Changes and Impact:} Methods added to handle login, waste management data analysis, and performance monitoring.

\begin{itemize}
    \item \texttt{login(username: string, password: string): boolean} - Handles the login process and returns a boolean indicating success.
    \item \texttt{viewAndAnalyzeWasteManagementData(): string} - Provides analysis of waste management data.
    \item \texttt{monitorRecyclingAndWasteManagementPerformance(): string} - Monitors performance metrics and provides analysis.
\end{itemize}

\paragraph{PaymentGateway Class}

\textbf{Key Changes and Impact:} Method added to manage transactions and payments.

\begin{itemize}
    \item \texttt{manageTransactionsAndPayments(): void} - Manages all transactions and payments.
\end{itemize}

\subsubsection{Analysis of Method Parameters in the Enhanced Class Diagram}

The integration of method parameters within the class diagram in our system design was meticulously analyzed to demonstrate their alignment with the respective class attributes, thereby illuminating the intrinsic functionalities of the system. Here, we elaborate on several key methods to illustrate how method parameters encapsulate system operations and enhance the overall clarity of the design documentation.

\paragraph{User Class Methods}
\begin{itemize}
    \item \textbf{registerUser(name: string, email: string, phoneNumber: string, address: string): boolean} - This method ensures comprehensive data collection essential for user registration by directly mapping each parameter to an existing class attribute. It is fundamental for maintaining data integrity and managing user information effectively.
    \item \textbf{listProduct(productDetails: ProductDetails): boolean} - The parameter \textit{productDetails} indicates a structured interaction with the \textit{ProductDetails} class, facilitating user-driven product management within the marketplace
    \item \textbf{trackWasteJourney(trackingCode: string): string} - Utilizing \textit{trackingCode}, this method provides critical functionality for tracking the disposal or lifecycle of products, enhancing transparency in waste management processes.
    \item \textbf{submitFeedbackOrReport(feedbackDetails: FeedbackDetails): boolean} - The structured input \textit{feedbackDetails} supports a detailed feedback mechanism, pivotal for system improvements and user satisfaction monitoring.
    \item \textbf{updateProfile(profileDetails: ProfileDetails): boolean} - This method, vital for user data accuracy and currency, allows for detailed modifications within user profiles using \textit{profileDetails}.
    \item \textbf{manageShippingAndDeliveryDetails(transactionID: string, shippingDetails: ShippingDetails): boolean} - Integrating \textit{transactionID} and \textit{shippingDetails}, this method underscores the complex logistics management associated with user transactions.
\end{itemize}

\paragraph{Transaction Class Methods}
\begin{itemize}
    \item \textbf{processTransaction(transactionDetails: TransactionDetails): boolean} - Here, \textit{transactionDetails} encompasses several critical attributes such as quantity, total price, and product identifiers, ensuring efficient and secure transaction processing.
\end{itemize}

\paragraph{Service and Payment Infrastructure}
\begin{itemize}
    \item The \textit{ServiceRequest} and \textit{PaymentGateway} classes illustrate the system’s infrastructure capable of handling detailed and complex service requests and financial transactions, respectively.
\end{itemize}

The detailed parameter analysis not only enhances the understanding of the system’s operations but also underscores the importance of precise and thoughtful parameter documentation in UML diagrams. This approach significantly contributes to the transparency and functionality of the system design, crucial for stakeholder comprehension and system maintenance.

\subsubsection{Method Validation}

The following table \ref{tab:method-validation} presents a detailed mapping of each method to its corresponding use case, illustrating the thorough validation of the methods added to the enhanced class diagram:

\begin{longtable}{|p{4cm}|p{8.7cm}|}
    \caption{Method Validation Against Use Cases} \\
    \hline
    \textbf{Use Case} & \textbf{Method} \\
    \hline
    \endfirsthead
    \hline
    \textbf{Use Case} & \textbf{Method}\\
    \hline
    \endhead
    \hline
    \endfoot
    \hline
    \endlastfoot
    UC1: User Registration & \texttt{registerUser(name: string, email: string, phoneNumber: string, address: string): boolean}\\
    \hline
    UC2: Listing Recyclable Waste Products & \texttt{listProduct(productDetails: ProductDetails): boolean} \\
    \hline
    UC3: Buying Recyclable Waste Products & \texttt{processTransaction(transactionDetails: TransactionDetails): boolean}  \\
    \hline
    UC4: Selling Recyclable Waste Products & \texttt{processSale(saleDetails: SaleDetails): boolean} \\
    \hline
    UC5: Reviewing Products or Services & \texttt{submitReview(reviewDetails: ReviewDetails): boolean} \\
    \hline
    UC7: Requesting Waste Collection Services & \texttt{submitCollectionRequest(requestDetails: CollectionRequestDetails): boolean} \\
    \hline
    UC8: Requesting Waste Transport Services & \texttt{submitTransportRequest(requestDetails: TransportRequestDetails): boolean} \\
    \hline
    UC11: Managing User Rewards & \texttt{manageRewards(): string}\\
    \hline
    UC12: Tracking Waste Journey & \texttt{trackWasteJourney(trackingCode: string): string} \\
    \hline
    UC13: Submitting Feedback or Reports & \texttt{submitFeedbackOrReport(feedbackDetails: FeedbackDetails): boolean}  \\
    \hline
    UC14: Viewing Service Requests and Status & \texttt{getServiceRequestDetails(requestID: string): string}  \\
    \hline
    UC15: Managing Platform Services & \texttt{login(username: string, password: string): boolean} \\
    \hline
    UC16: Managing User Profiles and Accounts & \texttt{updateProfile(profileDetails: ProfileDetails): boolean}  \\
    \hline
    UC17: Managing Transactions and Payments & \texttt{manageTransactionsAndPayments(): void} \\
    \hline
    UC18: Connecting with Transport Companies & \texttt{connectWithTransportCompanies(): string}  \\
    \hline
    UC19: Viewing and Analyzing Waste Management Data & \texttt{viewAndAnalyzeWasteManagementData(): string}  \\
    \hline
    UC22: Monitoring Recycling and Waste Management Performance & \texttt{monitorRecyclingAndWasteManagementPerformance(): string}  \\
    \hline
    UC23: Managing Shipping and Delivery Details & \texttt{manageShippingAndDeliveryDetails(transactionID: string, shippingDetails: ShippingDetails): boolean}  \\
    \label{tab:method-validation}
\end{longtable}

Incorporating ChatGPT's insights to enhance class diagrams with methods tailored to specific use cases demonstrates AI's usefulness in software design. The validation of each method, such as \texttt{registerUser} and \texttt{manageShippingAndDeliveryDetails}, against 20 distinct use cases confirms their relevance and specificity. This approach not only showcases the effectiveness of AI in iterative software development processes but also enhances system integrity and usability by ensuring that every component is tested for its intended purpose.

\subsubsection{Illustrative Examples of Enhanced Relationships}

The enhanced relationships within the class diagram are crucial for illustrating how the system components interact more effectively post-enhancement. Here are a few key examples:

\begin{itemize}
    \item \textbf{Inclusion of PlatformManager Class:} The newly introduced \textit{PlatformManager} class, featuring methods such as \textit{login}, \textit{viewAndAnalyzeWasteManagementData}, and \textit{monitorRecyclingAndWasteManagementPerformance}, embodies essential administrative functionalities for platform management. These methods signify the platform manager's responsibility for \textit{overseeing and analyzing waste management data and performance}, aligning with use cases UC19 and UC22.

    \item \textbf{User and CollectionPoint/RecyclingPoint Classes:} Significant improvements include direct connections between the \textit{User} class and the \textit{CollectionPoint} and \textit{RecyclingPoint} classes, highlighted in use cases UC7 (\textit{Requesting Waste Collection Services}) and UC9 (\textit{Viewing Collection and Recycling Points}). These connections clarify essential functionalities and interaction points for user interactions with waste management services.

    \item \textbf{Transaction and PaymentGateway Classes:} The enhanced relationship between \textit{PaymentGateway} and \textit{Transaction} with the \texttt{processTransaction()} method implies an active role for the payment gateway in transaction processing, in accordance with use cases such as UC3 (\textit{Buying Recyclable Waste Products}) and UC17 (\textit{Managing Transactions and Payments}).

    \item \textbf{WasteManagementService and CollectionPoint/RecyclingPoint Classes:} The newly established link between \textit{WasteManagementService} and the \textit{CollectionPoint} and \textit{RecyclingPoint} classes underscore its critical reliance on these entities for effective waste management, aligning with use cases UC1 and UC10 for service registration and waste management.
\end{itemize}

\subsubsection{New and Modified Relationships}

New and modified relationships were added to the class diagram based on the detailed use case analysis. This section presents the relationships observed after the enhancements (Table \ref{tab:new-modified-relationships}).

\begin{longtable}{|p{2cm}|p{3cm}|p{7cm}|}
\caption{New and Modified Relationships in the Class Diagram}
\label{tab:new-modified-relationships} \\
\hline
\textbf{Entities} & \textbf{Original Relationship} & \textbf{New/Modified Relationship} \\
\hline
\endfirsthead

\hline
\textbf{Entities} & \textbf{Original Relationship} & \textbf{New/Modified Relationship} \\
\hline
\endhead

\hline
\endfoot

\textbf{User and Product} & 
User "1" -- "0..*" Product : hasProduct & 
User interactions with products have been enhanced with methods like \texttt{listProduct(productDetails: ProductDetails): boolean}. \\
\hline
\textbf{User and Transaction} & 
User "1" -- "0..*" Transaction : hasTransaction & 
Methods like \newline \texttt{manageTransactionsAndPayments()} in the User class and \texttt{processTransaction(transactionDetails: TransactionDetails): boolean} in the Transaction class have been added. \\
\hline
\textbf{User and Review} & 
User "1" -- "0..*" Review : hasReview & 
The Review class now includes the method \texttt{submitReview(reviewDetails: ReviewDetails): boolean}. \\
\hline
\textbf{User and ServiceRequest} & 
Implicit through use cases & 
The ServiceRequest class includes methods such as \texttt{submitCollectionRequest(requestDetails: CollectionRequestDetails): boolean}. \\
\hline
\textbf{Transaction and PaymentGateway} & 
PaymentGateway "1" -- "0..*"

 Transaction : managesPayment & 
The PaymentGateway class now includes the method \texttt{manageTransactionsAndPayments()}. \\
\hline
\end{longtable}

\subsubsection{Relationship Validation}

To ensure that the new and modified relationships in the class diagram accurately represent the system's functionality, we validated them against the detailed use cases. Table \ref{tab:relationship-validation} presents this validation, showing how each new or modified relationship maps to the corresponding use cases.

\begin{longtable}{|p{7.5cm}|p{5cm}|}
\caption{Validation of New and Modified Relationships Against Use Cases}
\label{tab:relationship-validation} \\
\hline
\textbf{New/Modified Relationship} & \textbf{Referenced in Use Case(s)} \\
\hline
\endfirsthead

\hline
\textbf{New/Modified Relationship} & \textbf{Referenced in Use Case(s)} \\
\hline
\endhead

\hline
\endfoot

User interactions with products have been enhanced with methods like \texttt{listProduct(productDetails: ProductDetails): boolean} & 
UC2: Listing Recyclable Waste Products \newline
UC3: Buying Recyclable Waste Products \newline
UC4: Selling Recyclable Waste Products \\
\hline

Methods like \texttt{manageTransactionsAndPayments()} in the User class and \texttt{processTransaction(transactionDetails: TransactionDetails): boolean} in the Transaction class have been added & 
UC3: Buying Recyclable Waste Products \newline
UC4: Selling Recyclable Waste Products \newline
UC17: Managing Transactions and Payments \\
\hline

The Review class now includes the method \texttt{submitReview(reviewDetails: ReviewDetails): boolean} & 
UC5: Reviewing Products or Services \\
\hline

The ServiceRequest class includes methods such as \texttt{submitCollectionRequest(requestDetails: CollectionRequestDetails): boolean} & 
UC7: Requesting Waste Collection Services \newline
UC8: Requesting Waste Transport Services \\
\hline

The PaymentGateway class now includes the method \texttt{manageTransactionsAndPayments()} & 
UC3: Buying Recyclable Waste Products \newline
UC4: Selling Recyclable Waste Products \newline
UC17: Managing Transactions and Payments \\
\hline
\end{longtable}

This validation ensures that the system's functionality aligns with the user requirements and the platform's operational goals. The relationships such as enhanced product interactions, transaction and payment management, review submissions, and service requests have been effectively mapped to their corresponding use cases, confirming their necessity and implementation correctness. These validated relationships confirm that the class diagram accurately represents the system's necessary interactions and supports the comprehensive functionality required by the platform. This alignment between the class diagram and use cases ensures that the design is robust, user-centric, and prepared to handle real-world operations effectively.

\subsection{Classes Without Methods}

The following classes do not contain any methods and were not initially covered in the documentation:
\begin{itemize}
    \item Product
    \item CollectionPoint
    \item RecyclingPoint
    \item EnvironmentalImpact
    \item InformationResource
    \item RewardSystem
    \item PaymentDetails
    \item ShippingDetails
    \item TransportCompany
    \item StateAdministration
\end{itemize}

Additionally, subclasses such as IndividualUser and CorporateUser (extending the User class) and Plastic, Paper, Metal, and Glass (extending the Product class) also do not have any specific methods defined for them.

Future iterations should focus on integrating these unimplemented methods and refining class roles to optimize system functionality and user interaction, ensuring a robust and comprehensive software architecture.

\subsection{Conclusion}

The results section has provided a comprehensive presentation of the enhanced class diagrams, the methods added to key classes, and the new and modified relationships among classes. These enhancements, guided by the functional requirements extracted through the use of ChatGPT to analyze natural language use cases, demonstrate the methodology's effectiveness. Further discussions and interpretations of these results will be presented in the subsequent "Discussion" section.

\newpage

\section{Discussion}

Our study presents a novel AI-assisted methodology to enhance UML class diagrams by integrating NLP capabilities through ChatGPT. Unlike traditional methodologies that rely on manual translation of use case scenarios into design elements, our approach leverages AI to automate and refine this process, leading to more accurate and dynamic class diagrams.

\subsection{Key Findings and Methodological Advancements}

Traditional methodologies for enhancing UML class diagrams often involve substantial manual effort and are susceptible to errors and inconsistencies, such as:
\begin{itemize}
    \item \textbf{Errors in Translation:} Misinterpretation of use case scenarios leading to incorrect or incomplete design elements.
    \item \textbf{Inconsistencies:} Variations in how different team members interpret and document use cases.
    \item \textbf{Time-Consuming Processes:} Manual processes are slow and require significant effort from skilled developers.
\end{itemize}

To address these challenges, our study introduces a novel AI-assisted approach leveraging ChatGPT to automate the extraction of methods from natural language use case tables and integrate them into UML class diagrams. This advancement significantly enhances system understanding and streamlines the software development process. The key methodological advancements include:
\begin{enumerate}
    \item \textbf{Automated Method Extraction:} ChatGPT analyzes natural language use case tables to identify potential methods and their relationships, reducing the manual effort required for this task.
    \item \textbf{Dynamic Diagram Enhancement:} The extracted methods are iteratively integrated into the UML class diagrams, transforming them from static to dynamic representations that accurately reflect system behaviors.
    \item \textbf{Improved Accuracy and Consistency:} The automation process minimizes human error and ensures consistent application of design elements across the entire diagram.
    \item \textbf{Time Efficiency:} By automating the extraction and integration process, the time and effort required to enhance class diagrams are significantly reduced.
\end{enumerate}

These advancements showcase the significant benefits of integrating AI with traditional software design processes, making the overall design process more efficient and less prone to error.

\subsection{Interpretation of Results and Practical Implications}

Integrating ChatGPT to enhance UML class diagrams led to substantial improvements in system design accuracy and comprehensiveness. Initially, the class diagram was a static representation of the system's structure. Through iterative applications of ChatGPT, the diagram was enriched by adding methods and refining relationships, capturing the full spectrum of system behaviors outlined in the use cases.

\subsubsection{Practical Implications:}
\begin{itemize}
    \item \textit{Accuracy and Comprehensiveness:} ChatGPT identified and incorporated methods from natural language use cases, ensuring the diagram accurately reflected system functionalities and served as a more complete blueprint for implementation.
    \item \textit{Dynamic Method Integration:} The iterative use of ChatGPT allowed continuous refinement of the class diagrams. Each method added was evaluated for alignment with system requirements, ensuring relevance and accuracy.
    \item \textit{System Behavior Representation:} Adding methods bridged the gap between structural design and system behavior, providing a dynamic representation of functionalities.
    \item \textit{Broad Implications for Software Engineering:} Beyond our project, ChatGPT's capabilities have significant implications for software engineering. It can generate documentation, suggest code improvements, and facilitate real-time collaboration among development teams. However, challenges such as ensuring the contextual accuracy of extracted methods and the need for continuous training to adapt to domain-specific language remain. Addressing these challenges is crucial for maximizing the effectiveness of AI tools in software development.
\end{itemize}

\subsubsection{Examples of Enhancements:}
\begin{itemize}
    \item \textit{Transforming Static Diagrams:} By interpreting natural language use cases, ChatGPT autonomously transforms static class diagrams into dynamic models reflecting system functionalities. For instance, methods such as \texttt{login}, \texttt{monitorRecyclingAndWasteManagementPerformance}, and \texttt{viewAndAnalyzeWasteManagementData} were added to relevant classes like \texttt{User}, \texttt{Transaction}, and \texttt{PlatformManager}, resulting in more comprehensive and functional diagrams.
    \item \textit{Intuitive Augmentations:} ChatGPT's advanced analytical capabilities were demonstrated when it autonomously added the \texttt{PlatformManager} class and established new relationships based on use case requirements, without explicit instructions. This included the addition of methods like \texttt{login}, \texttt{viewAndAnalyzeWasteManagementData}, and \texttt{monitorRecyclingAndWasteManagementPerformance}, enriching the diagram’s detail and accuracy.
\end{itemize}
\subsection{Agile Adaptability}

Our AI-driven approach enhances flexibility and responsiveness in Agile software development by dynamically updating class diagrams based on new use cases and requirements. This ensures the diagrams remain relevant throughout the development process, supporting continuous improvement and rapid iterations, key principles of Agile methodologies like Scrum \cite{schwaber2002} and Kanban \cite{kniberg2010}.

\subsubsection{Practical Benefits:}
\begin{itemize}
    \item \textbf{Flexibility and Responsiveness:} Dynamic updates keep class diagrams accurate and adaptable to changes without disrupting development.
    \item \textbf{Enhanced Clarity and Usability:} Updated diagrams improve traceability and alignment with system requirements, providing a comprehensive view of class interactions and relationships.
    \item \textbf{Support for Agile Principles:} Continuously refined diagrams align with evolving use cases, supporting iterative development, continuous feedback, and rapid adaptation.
\end{itemize}

\subsection{Method Implementation Based on Use Case Availability}

Implementing class methods based on use case availability provides a structured, efficient, and user-focused approach to software development. This ensures that each class in the system is functional, maintainable, and aligned with real-world requirements, leading to a more robust and user-friendly application.

\subsubsection{Key Aspects:}
\begin{enumerate}
    \item \textbf{Targeted Development:}
    \begin{itemize}
        \item \textit{Efficiency:} Focuses development efforts on explicitly required functionalities, avoiding unnecessary code.
        \item \textit{Relevance:} Ensures methods implemented are directly relevant to the application's needs.
    \end{itemize}
    \item \textbf{Improved Maintainability:}
    \begin{itemize}
        \item \textit{Clear Requirements:} Methods are added based on well-documented use cases, making the code easier to understand and maintain.
        \item \textit{Simplified Updates:} Changes in requirements can be addressed by updating specific, well-documented methods, reducing modification complexity.
    \end{itemize}
    \item \textbf{User-Centric Development:}
    \begin{itemize}
        \item \textit{User Needs Focus:} Aligns development with user needs and requirements, enhancing user satisfaction.
        \item \textit{Feedback Incorporation:} Facilitates incorporating user feedback into specific, actionable changes.
    \end{itemize}
    \item \textbf{Consistency Across Classes:}
    \begin{itemize}
        \item \textit{Uniform Enhancements:} Ensures all classes are enhanced consistently based on the availability of use cases, maintaining uniform functionality across the system.
        \item \textit{System Integrity:} Maintains overall system integrity by ensuring each class has methods that are necessary and sufficient for its role.
    \end{itemize}
    \item \textbf{Scalability and Extensibility:}
    \begin{itemize}
        \item \textit{Modular Growth:} Allows for the addition of new methods as new use cases arise, enabling controlled and modular system growth.
        \item \textit{Adaptability:} Adapts to new requirements and use cases without extensive refactoring, making the system more scalable and extensible.
    \end{itemize}
\end{enumerate}

\subsection{Implications for Software Engineering}

Integrating AI, specifically NLP, into software modeling tasks can significantly enhance efficiency and accuracy by automating routine tasks and ensuring dynamic updates. This reduces the workload on developers, allowing them to focus on crucial design decisions and innovative solutions.

\subsubsection{Benefits:}
\begin{itemize}
\item \textbf{Enhanced Efficiency and Accuracy:} Automating repetitive tasks allows developers to focus on complex design decisions, with continuous updates to class diagrams ensuring they reflect the latest requirements and use cases accurately.
\item \textbf{Reduced Developer Workload:} Automation frees developers from routine tasks, allowing more time for critical design aspects, thus accelerating the development process by quickly translating use cases into class diagrams.
\item \textbf{Improved Software Model Quality:} Clearer, more maintainable class diagrams improve understanding and communication among team members and stakeholders. Automation minimizes human error, ensuring accurate and consistent diagrams.
\item \textbf{Agile Compatibility:} AI-driven approaches support Agile methodologies by facilitating continuous feedback and rapid iterations, keeping design aligned with evolving requirements.
\item \textbf{Innovation and Adaptability:} Automated updates to class diagrams enable quick adaptation to changing requirements or new technologies, fostering exploration of alternative designs.
\item \textbf{Adherence to Standards:} Ensures diagrams conform to UML best practices and naming conventions.
\end{itemize}
\subsubsection{Challenges:}
\begin{itemize}
    \item \textbf{Human Oversight:} Despite the automation capabilities, human judgment remains crucial for overseeing complex decisions and ensuring the integrity of design refinements.
    \item \textbf{Bias Mitigation:} Addressing inherent biases in AI algorithms is essential to maintain fairness and objectivity in the design process.
\end{itemize}

This study illustrates the transformative potential of AI in software engineering, paving the way for more innovative, efficient, and adaptive methodologies.

\subsection{Ethical Considerations}

The integration of AI technologies like ChatGPT in software engineering raises several ethical considerations. These include concerns about data privacy, as the training of such models often requires access to extensive datasets that may contain sensitive information. Additionally, there is the potential for dependency on automated systems, which could lead to a devaluation of human expertise and oversight in critical design processes. Ensuring transparent AI training processes and adhering to ethical standards and regulations is paramount to foster trust and accountability in automated systems.

\subsection{Detailed Analyses}

\subsubsection{Integration of Methods According to Use Cases}

The methodology presented in this study effectively integrates methods into a previously static class diagram, based on the detailed analysis of use case tables for a Waste Recycling Platform. Each step in the methodology contributes to the enrichment of the class diagram, ensuring that it reflects the functional requirements of the system.

The methods incorporated into the class diagram are directly derived from the actions and scenarios described in the use cases. For instance, the \texttt{User} class includes methods like \texttt{registerUser}, \texttt{listProduct}, and \texttt{submitFeedbackOrReport}, which align with use cases such as UC1 (Registration), UC2 (Listing Recyclable Waste), and UC8 (Submitting Feedback), respectively. Each method's parameters are meticulously chosen to match the required inputs for these actions, ensuring a seamless integration of functional requirements.

The Transaction class methods, such as \texttt{processTransaction} and \texttt{processSale}, correspond to use cases involving buying and selling recyclable waste (UC3 and UC4). These methods encapsulate transaction details, reflecting the complex interactions between users and the system during these processes.

The iterative process not only adds methods but also refines relationships between classes. For example, the enhanced relationship between \texttt{User} and \texttt{Product} classes, facilitated by methods like \texttt{listProduct}, illustrates the dynamic interaction required for product management. Similarly, the introduction of the \texttt{PlatformManager} class, with methods for logging in, analyzing waste management data, and monitoring performance, highlights the administrative functionalities necessary for platform management.

The study also identifies and integrates new relationships, such as those between \texttt{Transaction} and \texttt{PaymentGateway} classes, ensuring that payment processes are accurately represented. The validation of these relationships against the use cases confirms their relevance and correctness, enhancing the diagram's fidelity to the system's intended behavior.

The enhanced class diagram's behavioral alignment with the use cases ensures that each method added to the classes supports the system's functional requirements. The validation process, detailed in the results section, demonstrates that the methods are not only necessary but also correctly implemented. This validation is critical for maintaining the integrity and usability of the class diagram, providing stakeholders with a reliable blueprint for system implementation.

The validation against 23 distinct use cases ensures comprehensive coverage of the system's functionality, confirming that the enriched class diagram accurately represents both static structure and dynamic behavior. This alignment is crucial for effective communication, design, and implementation, reducing the risk of errors and inconsistencies in the final system.

\subsubsection{Unimplemented Methods}

The following table (\ref{tab:not-implemented-methods}) provides examples of methods that were not implemented for specific use cases. This highlights the gaps that need to be addressed to ensure comprehensive system functionality.

\begin{table}[h]
\centering
\begin{tabular}{|p{3cm}|p{5cm}|p{4cm}|}
\hline
\textbf{Use Case} & \textbf{Proposed Implementation} \\
\hline
UC6: Accessing Educational Resources & Implement \texttt{accessResources()} method in \texttt{InformationResource} class. \\
\hline
UC9: Viewing Collection and Recycling Points & Implement \texttt{viewCollectionPoints()} and \texttt{viewRecyclingPoints()} methods in \texttt{CollectionPoint} and \texttt{RecyclingPoint} classes. \\
\hline
UC10: Monitoring Environmental Impact & Implement \texttt{viewEnvironmentalImpact(productID: string)} method in \texttt{EnvironmentalImpact} class. \\
\hline
\end{tabular}
\caption{Not Implemented Methods}
\label{tab:not-implemented-methods}
\end{table}

\paragraph{Potential Reasons for Omitted Methods}
\begin{itemize}
    \item \textbf{UC6:} Involves information retrieval, which ChatGPT might have deprioritized.
    \item \textbf{UC9:} Viewing data might have been seen as less dynamic.
    \item \textbf{UC10:} Monitoring and analyzing data could have been overlooked due to the focus on action-driven methods.
\end{itemize}

Implementing class methods based on use case availability provides a structured, efficient, and user-focused approach to software development. This ensures that each class in the system is functional, maintainable, and aligned with real-world requirements, leading to a more robust and user-friendly application. For example, while methods for listing, buying, and selling products were added to related classes (User, Transaction), the Product class lacks methods due to the absence of explicit use cases detailing necessary operations for products.

\subsection{Comparison with Other Works}

This subsection compares our AI-assisted methodology for enhancing UML class diagrams with other approaches in terms of accuracy, efficiency, time, and alignment with Agile practices.

\subsubsection{Accuracy}
Our methodology leverages detailed use case tables and advanced NLP capabilities of ChatGPT to ensure high accuracy in extracting and integrating methods. This results in more comprehensive and functional UML diagrams compared to other methods. For instance, Egyed's abstracted UML models \cite{Egyed2002} and Eichelberger's improved layouts \cite{Eichelberger2002} lack the depth and accuracy in method extraction. Similarly, while Nasiri \cite{Nasiri2020}, Sharma \cite{Sharma2010}, Lucassen \cite{Lucassen2017}, and Kumar \cite{Kumar2008} aim for accuracy, they do not match the comprehensiveness of our approach. Anda and Sjøberg \cite{Anda2005} generate enhanced UML diagrams but fall short in method extraction and integration accuracy.

\subsubsection{Efficiency}
Our approach enhances efficiency by automating method extraction and integration, reducing manual effort and errors. Unlike traditional methods that require manual translation of use case scenarios \cite{Egyed2002, Eichelberger2002}, our use of AI provides a distinct advantage. Nasiri \cite{Nasiri2020} and Sharma \cite{Sharma2010} improve efficiency with structured requirements, but our approach further minimizes human intervention, enhancing overall efficiency.

\subsubsection{Time}
The iterative and real-time updating capabilities of our methodology significantly reduce the time required to generate UML diagrams. This is a substantial improvement over the more time-consuming manual processes used by Egyed and Eichelberger \cite{Egyed2002, Eichelberger2002}. While Nasiri \cite{Nasiri2020} and Sharma \cite{Sharma2010} offer moderate support for iterative processes, our real-time updating capability ensures immediate integration of changes, making it more time-efficient.

\subsubsection{Alignment with Agile Practices}
Our methodology supports Agile practices through iterative development, continuous feedback, and rapid adaptation. The real-time updating and iterative enhancement capabilities ensure class diagrams remain relevant, aligning with Agile principles \cite{schwaber2002, kniberg2010}. Egyed \cite{Egyed2002} and Osman \cite{Osman2014} provide limited Agile support, while Nasiri \cite{Nasiri2020} and Sharma \cite{Sharma2010} offer moderate alignment. Our strong alignment with Agile methodologies enhances applicability in modern software development.

\paragraph{Automated Method Extraction}
Previous tools like those described in \cite{Abdelnabi2021, Briand2012, Omer2021} lack the capability to automatically extract and integrate methods, resulting in incomplete diagrams. Our methodology automates method extraction, reducing manual effort and enhancing diagram completeness.

\paragraph{Dynamic Enhancement}
Tools such as CM-Builder \cite{Omer2021}, Automated Conceptual Model \cite{Zhao2021}, and REBUILDER UML \cite{Abdelnabi2021} generate static diagrams that require significant user intervention for updates. In contrast, our approach dynamically enhances class diagrams iteratively, reflecting the latest functional requirements and supporting agile development practices.

\paragraph{Reduced User Intervention}
High levels of user intervention required by tools such as the D-H project \cite{Omer2021}, LIDA tool \cite{P2020}, and semi-automated approaches \cite{Perez2021} are a common limitation. Our AI-driven methodology minimizes manual refinement, making the process more efficient and less error-prone, which is especially beneficial in large-scale projects.

\paragraph{Simplicity of Use}
The simplicity of use is another significant advantage. Tools like MOVA \cite{Abdelnabi2021} and RACE \cite{Maatuk2021} are complex to set up and require extensive manual intervention. Our AI-driven methodology automates complex tasks, making the process user-friendly and accessible, even for those with limited technical expertise.

\subsubsection{Emphasis on System Behavior/Dynamism}
Our methodology excels in capturing the system's dynamic behavior through automated method extraction and iterative diagram enhancement. This focus on behavior is a significant advancement over previous tools. By continuously updating the diagrams to reflect new methods and relationships, our approach ensures structurally sound and behaviorally comprehensive models, providing a robust blueprint for system implementation.

By addressing the limitations of existing tools—such as static diagrams, high user intervention, and limited method extraction—our approach offers a more dynamic, efficient, and comprehensive solution for class diagram generation.

\subsection{Limitations and Threats to Validity}

While our approach offers significant improvements, it also presents some limitations and potential threats to validity that need consideration.

\begin{enumerate}
    \item \textbf{Input Quality Dependence}:
        \begin{itemize}
            \item \textit{Limitation}: The accuracy of the enhancements depends significantly on the quality and clarity of input use case tables. Poor documentation can lead to incorrect method extraction and integration.
            \item \textit{Mitigation}: Ensuring high-quality, detailed use case documentation and incorporating a validation step to check clarity before processing can mitigate this limitation.
        \end{itemize}
    \item \textbf{NLP Challenges}:
        \begin{itemize}
            \item \textit{Limitation}: ChatGPT's natural language processing capabilities might require human oversight to ensure accuracy, particularly with domain-specific nuances. Misinterpretations can occur, leading to inaccuracies in the generated diagrams.
            \item \textit{Mitigation}: Incorporating domain-specific training data for ChatGPT and including a review process by domain experts can address this issue.
        \end{itemize}
    \item \textbf{Single Use Case Limitation}:
        \begin{itemize}
            \item \textit{Limitation}: The study's reliance on a single use case may affect the generalizability of the methodology. Different systems and use cases might present unique challenges that this study did not address.
            \item \textit{Mitigation}: Validating the methodology across diverse use cases and larger systems can confirm its effectiveness and scalability.
        \end{itemize}
    \item \textbf{Human Oversight and AI Bias}:
        \begin{itemize}
            \item \textit{Limitation}: Inherent biases in AI algorithms can affect the interpretation of use cases, highlighting the importance of human intervention. AI bias can lead to skewed results and unfair or incorrect system design decisions.
            \item \textit{Mitigation}: Implementing rigorous testing and review protocols to identify and mitigate biases and ensuring continuous training and updates to the AI model to adapt to new data and requirements are essential steps.
        \end{itemize}
    \item \textbf{Scalability}:
        \begin{itemize}
            \item \textit{Limitation}: The methodology's performance and scalability when applied to large-scale projects or diverse domains have not been thoroughly tested.
            \item \textit{Mitigation}: Applying the AI-driven approach to large-scale projects in different domains and analyzing performance metrics can ensure its robustness and flexibility for different project sizes and complexities.
        \end{itemize}
    \item \textbf{Empirical Studies}:
        \begin{itemize}
            \item \textit{Limitation}: The lack of empirical studies measuring the impact of this methodology on development efficiency and project outcomes could limit its perceived validity.
            \item \textit{Mitigation}: Conducting empirical studies to compare AI-driven methodology with traditional methods and collecting and analyzing data on development efficiency, accuracy, and user satisfaction can provide quantitative and qualitative evidence of the methodology's benefits and challenges.
        \end{itemize}
\end{enumerate}

\subsection{Future Research Directions}

Future research will prioritize refining methodologies, expanding applications, and assessing the impact on design efficiency, maintenance, and system scalability in software engineering. The following prioritized areas represent the most critical aspects to be addressed:

\begin{enumerate}
    \item \textbf{Improving Use Case Documentation}:
        \begin{itemize}
            \item \textit{Objective}: Enhance the quality and clarity of use case documentation to improve AI-driven method extraction.
            \item \textit{Plan}: Develop standardized templates and guidelines for writing detailed and unambiguous use cases.
            \item \textit{Hypothesis}: Standardized and high-quality use case tables will significantly improve the accuracy of method extraction by ChatGPT, resulting in more precise and complete class diagrams.
            \item \textit{Connection to Findings}: This direction addresses the identified limitation of input quality dependence, ensuring that the AI-driven process can consistently produce accurate results.
        \end{itemize}
    \item \textbf{Enhancing NLP Capabilities}:
        \begin{itemize}
            \item \textit{Objective}: Integrate advanced NLP models to handle more complex use cases effectively.
            \item \textit{Plan}: Incorporate domain-specific training data to fine-tune ChatGPT for better comprehension of specialized terminology and context. Collaborate with NLP researchers to enhance the model's capabilities.
            \item \textit{Hypothesis}: Advanced and domain-specific NLP models will improve the accuracy and relevance of method extraction, especially for complex and niche software systems.
            \item \textit{Connection to Findings}: This direction aims to mitigate the NLP challenges identified, improving the system's ability to interpret and process domain-specific language accurately.
        \end{itemize}
    \item \textbf{Developing Automated Validation Tools}:
        \begin{itemize}
            \item \textit{Objective}: Create tools to automatically verify the correctness and completeness of class diagrams.
            \item \textit{Plan}: Develop validation algorithms that can compare generated diagrams against use case requirements and industry standards. Integrate these tools into the AI-driven process for real-time validation.
            \item \textit{Hypothesis}: Automated validation tools will reduce the need for manual reviews, ensuring the diagrams consistently meet specified requirements and standards.
            \item \textit{Connection to Findings}: This addresses the need for human oversight and reduces the dependency on manual validation, enhancing the overall reliability of the AI-generated diagrams.
        \end{itemize}
    \item \textbf{Testing Scalability}:
        \begin{itemize}
            \item \textit{Objective}: Evaluate the methodology's performance and scalability on larger, more complex systems.
            \item \textit{Plan}: Apply the AI-driven approach to large-scale projects in various domains, such as healthcare, finance, and manufacturing. Analyze performance metrics and identify scalability challenges.
            \item \textit{Hypothesis}: The methodology will be scalable and adaptable across different domains, demonstrating its robustness and flexibility for various project sizes and complexities.
            \item \textit{Connection to Findings}: This directly addresses the scalability limitation, ensuring that the methodology can be applied effectively to a wide range of applications.
        \end{itemize}
    \item \textbf{Conducting Empirical Studies}:
        \begin{itemize}
            \item \textit{Objective}: Measure the impact of the AI-driven methodology on development efficiency and project outcomes.
            \item \textit{Plan}: Design and implement empirical studies to compare AI-driven methodology with traditional methods. Collect and analyze data on development efficiency, accuracy, and user satisfaction.
            \item \textit{Hypothesis}: Empirical studies will provide quantitative and qualitative evidence of the benefits and challenges of using AI-driven methodologies, guiding future improvements and adoption.
            \item \textit{Connection to Findings}: This addresses the limitation of lacking empirical validation, providing concrete data to support the efficacy of the AI-driven approach.
        \end{itemize}
    \item \textbf{Addressing Ethical and Bias Considerations}:
        \begin{itemize}
            \item \textit{Objective}: Ensure ethical use and mitigate biases in automated diagram generation.
            \item \textit{Plan}: Implement protocols for ethical AI use and continuous monitoring to identify and address biases in AI algorithms.
            \item \textit{Hypothesis}: Ethical guidelines and bias mitigation strategies will enhance the fairness and objectivity of AI-driven processes.
            \item \textit{Connection to Findings}: This responds to the identified need for human oversight and AI bias mitigation, ensuring responsible AI implementation.
        \end{itemize}
    \item \textbf{Optimizing Human-AI Collaboration}:
        \begin{itemize}
            \item \textit{Objective}: Integrate human oversight with AI-driven methodologies for optimal performance.
            \item \textit{Plan}: Develop frameworks for effective human-AI collaboration, leveraging the strengths of both for improved outcomes.
            \item \textit{Hypothesis}: Balanced human-AI collaboration will enhance the quality and reliability of class diagram generation.
            \item \textit{Connection to Findings}: This addresses the need for human intervention in complex decision-making, ensuring that AI complements rather than replaces human expertise.
        \end{itemize}
\end{enumerate}

By clearly prioritizing these future research directions and tying each to specific findings and limitations of the current study, we ensure a focused and actionable path forward for improving and expanding the AI-driven methodology.

\subsection{Responses to Research Questions}

\begin{itemize}
    \item \textbf{RQ1: To what extent is it feasible to integrate ChatGPT to dynamically enhance the class diagram dynamics effectively?}
        \begin{itemize}
            \item \textit{Response}: The integration of ChatGPT into the process of enhancing UML class diagrams has proven to be highly feasible. ChatGPT successfully analyzed natural language use case tables to identify relevant methods for each class, dynamically updating the initial static diagram. This iterative enrichment process, guided by AI, ensured that the class diagram evolved to accurately represent the system's functional requirements.
        \end{itemize}
    \item \textbf{RQ2: How does the AI-driven enhancement of UML class diagrams impact the efficiency and accuracy of software modeling compared to traditional manual methods?}
        \begin{itemize}
            \item \textit{Response}: AI-driven enhancement of UML class diagrams significantly improves both the efficiency and accuracy of software modeling. By automating the extraction and integration of methods from detailed use cases, ChatGPT reduced the time and effort required for these tasks, enhancing the efficiency of the modeling process. Additionally, the AI-driven approach minimized human error and ensured consistent application of design elements, leading to more accurate and comprehensive class diagrams.
        \end{itemize}
    \item \textbf{RQ3: What are the limitations and challenges of using ChatGPT for dynamic class diagram enhancement, and how can these be mitigated?}
        \begin{itemize}
            \item \textit{Response}: Using ChatGPT for dynamic class diagram enhancement presents several limitations and challenges. One major limitation is the AI's reliance on the quality and clarity of the initial use case scenarios, which can impact the accuracy of the enhancements. Additionally, ChatGPT may struggle with domain-specific terminology or ambiguous use case descriptions, potentially leading to inaccuracies. To mitigate these challenges, it is essential to ensure high-quality, detailed use case documentation and incorporate human oversight to verify the AI-generated enhancements. Future research should focus on improving the AI's comprehension capabilities and developing automated validation tools to ensure the integrity and accuracy of the enhanced class diagrams.
        \end{itemize}
\end{itemize}

\subsection{Conclusion}

In summary, utilizing ChatGPT to enhance class diagrams by analyzing natural language use cases has proven effective. The enhanced diagrams accurately reflect functional requirements, covering most use cases provided. This study highlights AI tools' potential to improve software design processes, contributing to both theoretical and practical advancements in the field. This study presents a novel approach to dynamically enhancing UML class diagrams by leveraging ChatGPT to analyze natural language use cases. The methodology significantly improves the clarity, maintainability, and functional completeness of the class diagrams, offering a robust blueprint for system implementation. Future research will focus on addressing identified limitations and expanding the methodology's applicability to broader and more complex systems.

\section{Conclusion}

This study explored the integration of ChatGPT, an advanced AI language model, into the process of dynamically enhancing UML class diagrams. By leveraging NLP techniques, we demonstrated that AI can effectively automate the extraction of methods and interactions from detailed use case tables and incorporate them into class diagrams. This approach not only improves the accuracy and completeness of the diagrams but also significantly reduces the manual effort required for their creation and maintenance.

Including methods in class diagrams provides a detailed specification of the functionalities a class offers, representing both the structure and behavior of the system. This practice enhances documentation, design clarity, and maintainability while supporting modularity and facilitating object-oriented design principles. The findings of this study indicate that AI-driven methodologies hold substantial potential for transforming traditional software engineering practices.

The study found that integrating ChatGPT significantly improved the accuracy and efficiency of class diagram creation and maintenance. The dynamic enhancement of class diagrams using ChatGPT provides a more comprehensive representation of system behaviors and interactions, which is crucial for accurate system modeling and design. This approach aligns well with Agile development practices, facilitating rapid iterations and continuous improvement.

However, the methodology's effectiveness is contingent on the quality of the input use case tables and may require human oversight to ensure accuracy. Future research should explore the application of AI-driven methodologies to other types of software models and address the identified limitations to further enhance their applicability.

In conclusion, the integration of AI, particularly NLP techniques like those employed by ChatGPT, represents a significant advancement in software engineering practices. By automating routine tasks and ensuring dynamic updates, AI-driven approaches can transform traditional methodologies, leading to more accurate, efficient, and innovative software development processes. Continued research and development in this area are imperative to harness the full capabilities of AI and advance the state of software engineering.

\section*{Declarations}

\begin{itemize}
\item \textbf{Funding:} The authors declare that they have no conflict of interest.
\item \textbf{Conflict of interest/Competing interests:} The authors declare that they have no conflict of interest.
\item \textbf{Ethics approval and consent to participate:} Not applicable. This study did not involve human participants, human data, or human tissue.
\item \textbf{Consent for publication:} All authors have given their consent for publication of this manuscript.
\item \textbf{Data availability:} \item \textbf{Data availability:} All data generated or analyzed during this study are included in this published article.
\item \textbf{Materials availability:} All materials utilized in this research are detailed within the article. No additional materials are available.
\item \textbf{Code availability:} Not applicable.
\item \textbf{Author contribution:} 
\begin{itemize}
    \item Djaber Rouabhia conceived and designed the study, conducted the experiments and wrote the manuscript.
    \item Ismail Hadjadj analyzed the data and reviewed the manuscript.
\end{itemize}
\end{itemize}

\bigskip
\begin{flushleft}%
Editorial Policies for:

\bigskip\noindent
Springer journals and proceedings: \url{https://www.springer.com/gp/editorial-policies}

\bigskip\noindent
Nature Portfolio journals: \url{https://www.nature.com/nature-research/editorial-policies}

\bigskip\noindent
\textit{Scientific Reports}: \url{https://www.nature.com/srep/journal-policies/editorial-policies}

\bigskip\noindent
BMC journals: \url{https://www.biomedcentral.com/getpublished/editorial-policies}
\end{flushleft}

\begin{appendices}
\section{Detailed Use Case Tables}\label{secA}
The purpose of this appendix is to provide a comprehensive reference of the 23 detailed use case tables that support the main article. These tables have been formulated in natural language and structured with columns such as Use Case ID, Actor, Description, Pre-condition, Trigger, Main scenario, Post-condition, and Exceptions. By including these detailed use case tables in a separate appendix, the clarity and flow of the main article are maintained, ensuring that readers can focus on the core content without being overwhelmed by extensive technical details. Each use case table offers a granular view of the interactions and functionalities required by the system, serving as a primary reference for identifying the necessary methods for each class to facilitate the described functionalities. The tables were originally derived by master’s degree students in computer science from a waste recycling platform specification document, underscoring their practical relevance and depth.

The following sections present the detailed use case tables for various user interactions and system functionalities, providing a robust foundation for understanding the dynamic behaviors within the system.
\subsection{UC1: User Registration}
\begin{longtable}{|>{\raggedright\arraybackslash}p{3.5cm}|>{\raggedright\arraybackslash}p{9.5cm}|}
\hline
\textbf{Attribute} & \textbf{Details} \\ \hline
\textbf{Actor} & Visitor \\ \hline
\textbf{Use Case} & User Registration \\ \hline
\textbf{Description} & A visitor to the platform registers as a user. \\ \hline
\textbf{Pre-conditions} & Visitor is not registered. \\ \hline
\textbf{Triggers} & Visitor selects to register \\ \hline
\textbf{Main Scenario} & 
\begin{enumerate}
    \item Visitor accesses registration page.
    \item Visitor provides required information (name, contact details, etc.).
    \item Visitor submits the registration form.
    \item System validates the information.
    \item System creates a new user account.
    \item Visitor receives confirmation of successful registration.
\end{enumerate} \\ \hline
\textbf{Post-conditions} & Visitor is registered and can log in. \\ \hline
\textbf{Extensions} & 
\begin{enumerate}
    \item Invalid information provided.
    \item User already exists.
    \item Technical issues during registration.
\end{enumerate} \\ \hline
\textbf{Relationships} & 
\begin{itemize}
    \item Visitor is associated with User Registration.
    \item User Registration extends to Authentication for subsequent logins.
\end{itemize} \\ \hline
\end{longtable}
\subsection{UC2: Listing Recyclable Waste Products}
\begin{longtable}{|>{\raggedright\arraybackslash}p{3.5cm}|>{\raggedright\arraybackslash}p{9.5cm}|}
\hline
\textbf{Attribute} & \textbf{Details} \\ \hline
\textbf{Actor} & User \\ \hline
\textbf{Use Case} & Listing Recyclable Waste Products \\ \hline
\textbf{Description} & User lists a recyclable waste product for sale. \\ \hline
\textbf{Pre-conditions} & User is registered and logged in. \\ \hline
\textbf{Triggers} & User chooses to list a product. \\ \hline
\textbf{Main Scenario} & 
\begin{enumerate}
    \item User accesses product listing page.
    \item User enters product details.
    \item User submits the listing.
    \item System validates and publishes the listing.
\end{enumerate} \\ \hline
\textbf{Post-conditions} & Product is listed on the platform. \\ \hline
\textbf{Extensions} & 
\begin{enumerate}
    \item Invalid product details.
    \item Duplicate product listing.
    \item Technical issues during listing.
\end{enumerate} \\ \hline
\textbf{Relationships} & Related to User Registration and Product Management. \\ \hline
\end{longtable}
\subsection{UC3: Buying Recyclable Waste Products}
\begin{longtable}{|>{\raggedright\arraybackslash}p{3.5cm}|>{\raggedright\arraybackslash}p{9.5cm}|}
\hline
\textbf{Attribute} & \textbf{Details} \\ \hline
\textbf{Actor} & User \\ \hline
\textbf{Use Case} & Buying Recyclable Waste Products \\ \hline
\textbf{Description} & User purchases a recyclable waste product. \\ \hline
\textbf{Pre-conditions} & User is registered and logged in, product is listed. \\ \hline
\textbf{Triggers} & User selects a product to buy. \\ \hline
\textbf{Main Scenario} & 
\begin{enumerate}
    \item User browses products.
    \item User selects a product.
    \item User completes the purchase process.
    \item Transaction is processed.
\end{enumerate} \\ \hline
\textbf{Post-conditions} & Product is purchased and removed from listings. \\ \hline
\textbf{Extensions} & 
\begin{enumerate}
    \item Payment failure.
    \item Product out of stock.
    \item Technical issues during purchase.
\end{enumerate} \\ \hline
\textbf{Relationships} & Related to Product Listing and Payment Processing. \\ \hline
\end{longtable}
\subsection{UC4: Selling Recyclable Waste Products}
\begin{longtable}{|>{\raggedright\arraybackslash}p{3.5cm}|>{\raggedright\arraybackslash}p{9.5cm}|}
\hline
\textbf{Attribute} & \textbf{Details} \\ \hline
\textbf{Actor} & User \\ \hline
\textbf{Use Case} & Selling Recyclable Waste Products \\ \hline
\textbf{Description} & User sells a recyclable waste product. \\ \hline
\textbf{Pre-conditions} & User is registered and logged in, product is listed. \\ \hline
\textbf{Triggers} & User chooses to sell a product. \\ \hline
\textbf{Main Scenario} & 
\begin{enumerate}
    \item User accesses product selling page.
    \item User enters sale details.
    \item User submits the sale.
    \item System processes the sale.
\end{enumerate} \\ \hline
\textbf{Post-conditions} & Product is sold and transaction is completed. \\ \hline
\textbf{Extensions} & 
\begin{enumerate}
    \item Payment failure.
    \item Buyer cancels the purchase.
    \item Technical issues during sale.
\end{enumerate} \\ \hline
\textbf{Relationships} & Related to Product Listing and Transaction Management. \\ \hline
\end{longtable}
\subsection{UC5: Reviewing Products or Services}
\begin{longtable}{|>{\raggedright\arraybackslash}p{3.5cm}|>{\raggedright\arraybackslash}p{9.5cm}|}
\hline
\textbf{Attribute} & \textbf{Details} \\ \hline
\textbf{Actor} & User \\ \hline
\textbf{Use Case} & Reviewing Products or Services \\ \hline
\textbf{Description} & User provides feedback on a purchased product or received service. \\ \hline
\textbf{Pre-conditions} & User has completed a purchase or received a service. \\ \hline
\textbf{Triggers} & User opts to write a review. \\ \hline
\textbf{Main Scenario} & 
\begin{enumerate}
    \item User accesses review page.
    \item User writes and submits the review.
    \item System publishes the review.
\end{enumerate} \\ \hline
\textbf{Post-conditions} & Review is available for other users to see. \\ \hline
\textbf{Extensions} & 
\begin{enumerate}
    \item Inappropriate content in the review.
    \item Technical issues during review submission.
\end{enumerate} \\ \hline
\textbf{Relationships} & Linked with Product Purchase and Service Utilization. \\ \hline
\end{longtable}
\subsection{UC6: Accessing Educational and Awareness Resources}
\begin{longtable}{|>{\raggedright\arraybackslash}p{3.5cm}|>{\raggedright\arraybackslash}p{9.5cm}|}
\hline
\textbf{Attribute} & \textbf{Details} \\ \hline
\textbf{Actor} & User \\ \hline
\textbf{Use Case} & Accessing Educational and Awareness Resources \\ \hline
\textbf{Description} & User accesses resources for education and awareness about waste management. \\ \hline
\textbf{Pre-conditions} & Resources are available on the platform. \\ \hline
\textbf{Triggers} & User navigates to the educational resources section. \\ \hline
\textbf{Main Scenario} & 
\begin{enumerate}
    \item User selects the resources section.
    \item User browses various resources.
    \item User reads or downloads the desired materials.
\end{enumerate} \\ \hline
\textbf{Post-conditions} & User gains knowledge or information. \\ \hline
\textbf{Extensions} & 
\begin{enumerate}
    \item Resource not available.
    \item Technical issues accessing resources.
\end{enumerate} \\ \hline
\textbf{Relationships} & Supports User Education and Engagement. \\ \hline
\end{longtable}
\subsection{UC7: Requesting Waste Collection Services}
\begin{longtable}{|>{\raggedright\arraybackslash}p{3.5cm}|>{\raggedright\arraybackslash}p{9.5cm}|}
\hline
\textbf{Attribute} & \textbf{Details} \\ \hline
\textbf{Actor} & User \\ \hline
\textbf{Use Case} & Requesting Waste Collection Services \\ \hline
\textbf{Description} & User requests a service for waste collection. \\ \hline
\textbf{Pre-conditions} & User is registered and has waste to be collected. \\ \hline
\textbf{Triggers} & User decides to request waste collection. \\ \hline
\textbf{Main Scenario} & 
\begin{enumerate}
    \item User accesses the service request section.
    \item User fills out and submits the collection request form.
    \item Service provider receives the request and schedules the collection.
\end{enumerate} \\ \hline
\textbf{Post-conditions} & Waste collection service is scheduled. \\ \hline
\textbf{Extensions} & 
\begin{enumerate}
    \item Incomplete request form.
    \item No service providers available.
    \item Technical issues during request submission.
\end{enumerate} \\ \hline
\textbf{Relationships} & Linked with User Registration and Service Management. \\ \hline
\end{longtable}
\subsection{UC8: Requesting Waste Transport Services}
\begin{longtable}{|>{\raggedright\arraybackslash}p{3.5cm}|>{\raggedright\arraybackslash}p{9.5cm}|}
\hline
\textbf{Attribute} & \textbf{Details} \\ \hline
\textbf{Actor} & User \\ \hline
\textbf{Use Case} & Requesting Waste Transport Services \\ \hline
\textbf{Description} & User requests a service for transporting waste. \\ \hline
\textbf{Pre-conditions} & User is registered and requires waste transport. \\ \hline
\textbf{Triggers} & User decides to request waste transport. \\ \hline
\textbf{Main Scenario} & 
\begin{enumerate}
    \item User accesses the transport service request section.
    \item User fills out and submits the transport request form.
    \item Transport provider receives the request and schedules the transport.
\end{enumerate} \\ \hline
\textbf{Post-conditions} & Waste transport service is scheduled. \\ \hline
\textbf{Extensions} & 
\begin{enumerate}
    \item Incomplete request form.
    \item No transport providers available.
    \item Technical issues during request submission.
\end{enumerate} \\ \hline
\textbf{Relationships} & Linked with User Registration and Transport Service Management. \\ \hline
\end{longtable}
\subsection{UC9: Viewing Collection and Recycling Points}
\begin{longtable}{|>{\raggedright\arraybackslash}p{3.5cm}|>{\raggedright\arraybackslash}p{9.5cm}|}
\hline
\textbf{Attribute} & \textbf{Details} \\ \hline
\textbf{Actor} & User \\ \hline
\textbf{Use Case} & Viewing Collection and Recycling Points \\ \hline
\textbf{Description} & User views the locations of collection and recycling points. \\ \hline
\textbf{Pre-conditions} & Collection and recycling points are registered on the platform. \\ \hline
\textbf{Triggers} & User wishes to locate a collection or recycling point. \\ \hline
\textbf{Main Scenario} & 
\begin{enumerate}
    \item User accesses the map or list of points.
    \item User browses or searches for specific locations.
    \item User views details of selected points.
\end{enumerate} \\ \hline
\textbf{Post-conditions} & User is informed about the locations. \\ \hline
\textbf{Extensions} & 
\begin{enumerate}
    \item No points in the user's area.
    \item Technical issues accessing the map or list.
\end{enumerate} \\ \hline
\textbf{Relationships} & Supports Waste Collection and Recycling Services. \\ \hline
\end{longtable}
\subsection{UC10: Monitoring Environmental Impact of Products}
\begin{longtable}{|>{\raggedright\arraybackslash}p{3.5cm}|>{\raggedright\arraybackslash}p{9.5cm}|}
\hline
\textbf{Attribute} & \textbf{Details} \\ \hline
\textbf{Actor} & User \\ \hline
\textbf{Use Case} & Monitoring Environmental Impact of Products \\ \hline
\textbf{Description} & User accesses information about the environmental impact of products. \\ \hline
\textbf{Pre-conditions} & Environmental impact data is available for products. \\ \hline
\textbf{Triggers} & User wishes to understand the environmental impact of a product. \\ \hline
\textbf{Main Scenario} & 
\begin{enumerate}
    \item User selects a product.
    \item User views the environmental impact details provided.
    \item User gains insights into the product's sustainability.
\end{enumerate} \\ \hline
\textbf{Post-conditions} & User is informed about the product's environmental impact. \\ \hline
\textbf{Extensions} & 
\begin{enumerate}
    \item No impact data available.
    \item Technical issues accessing the data.
\end{enumerate} \\ \hline
\textbf{Relationships} & Linked with Product Listing and Environmental Awareness. \\ \hline
\end{longtable}
\subsection{UC11: Managing User Rewards}
\begin{longtable}{|>{\raggedright\arraybackslash}p{3.5cm}|>{\raggedright\arraybackslash}p{9.5cm}|}
\hline
\textbf{Attribute} & \textbf{Details} \\ \hline
\textbf{Actor} & User \\ \hline
\textbf{Use Case} & Managing User Rewards \\ \hline
\textbf{Description} & User participates in and manages their rewards and incentives. \\ \hline
\textbf{Pre-conditions} & User is eligible for rewards. \\ \hline
\textbf{Triggers} & User engages in activities that accrue rewards. \\ \hline
\textbf{Main Scenario} & 
\begin{enumerate}
    \item User completes qualifying activities.
    \item User accumulates rewards points.
    \item User redeems points for rewards or incentives.
\end{enumerate} \\ \hline
\textbf{Post-conditions} & User receives rewards or incentives. \\ \hline
\textbf{Extensions} & 
\begin{enumerate}
    \item Dispute over reward points.
    \item Technical issues in rewards management.
\end{enumerate} \\ \hline
\textbf{Relationships} & Supports User Engagement and Loyalty Programs. \\ \hline
\end{longtable}
\subsection{UC12: Tracking Waste Journey}
\begin{longtable}{|>{\raggedright\arraybackslash}p{3.5cm}|>{\raggedright\arraybackslash}p{9.5cm}|}
\hline
\textbf{Attribute} & \textbf{Details} \\ \hline
\textbf{Actor} & User \\ \hline
\textbf{Use Case} & Tracking Waste Journey \\ \hline
\textbf{Description} & User tracks the journey of waste from collection to final treatment. \\ \hline
\textbf{Pre-conditions} & Waste has been collected and is in the process of being treated. \\ \hline
\textbf{Triggers} & User wants to know the status of their waste. \\ \hline
\textbf{Main Scenario} & 
\begin{enumerate}
    \item User accesses the waste tracking feature.
    \item User inputs the tracking code.
    \item System displays the current status and location of the waste.
\end{enumerate} \\ \hline
\textbf{Post-conditions} & User is informed about the waste journey status. \\ \hline
\textbf{Extensions} & 
\begin{enumerate}
    \item Invalid tracking code.
    \item Technical issues in tracking system.
\end{enumerate} \\ \hline
\textbf{Relationships} & Linked with Waste Collection and Transport Services. \\ \hline
\end{longtable}
\subsection{UC13: Submitting Feedback or Reports}
\begin{longtable}{|>{\raggedright\arraybackslash}p{3.5cm}|>{\raggedright\arraybackslash}p{9.5cm}|}
\hline
\textbf{Attribute} & \textbf{Details} \\ \hline
\textbf{Actor} & User \\ \hline
\textbf{Use Case} & Submitting Feedback or Reports \\ \hline
\textbf{Description} & User submits feedback or reports related to the platform's services. \\ \hline
\textbf{Pre-conditions} & User has used a service or wishes to provide feedback. \\ \hline
\textbf{Triggers} & User has feedback or a report to submit. \\ \hline
\textbf{Main Scenario} & 
\begin{enumerate}
    \item User accesses the feedback section.
    \item User fills out and submits the feedback form.
    \item Feedback is received and processed by the platform.
\end{enumerate} \\ \hline
\textbf{Post-conditions} & Feedback or report is submitted and acknowledged. \\ \hline
\textbf{Extensions} & 
\begin{enumerate}
    \item Incomplete feedback form.
    \item Technical issues during submission.
\end{enumerate} \\ \hline
\textbf{Relationships} & Supports Continuous Improvement and User Engagement. \\ \hline
\end{longtable}
\subsection{UC14: Viewing Service Requests and Status}
\begin{longtable}{|>{\raggedright\arraybackslash}p{3.5cm}|>{\raggedright\arraybackslash}p{9.5cm}|}
\hline
\textbf{Attribute} & \textbf{Details} \\ \hline
\textbf{Actor} & User \\ \hline
\textbf{Use Case} & Viewing Service Requests and Status \\ \hline
\textbf{Description} & User views and tracks the status of their service requests. \\ \hline
\textbf{Pre-conditions} & User has made one or more service requests. \\ \hline
\textbf{Triggers} & User wants to check the status of a service request. \\ \hline
\textbf{Main Scenario} & 
\begin{enumerate}
    \item User accesses their service request history.
    \item User selects a request to view details.
    \item System displays the current status and details of the request.
\end{enumerate} \\ \hline
\textbf{Post-conditions} & User is updated on the status of their service requests. \\ \hline
\textbf{Extensions} & 
\begin{enumerate}
    \item Service request not found.
    \item Technical issues accessing request details.
\end{enumerate} \\ \hline
\textbf{Relationships} & Linked with Service Request Management. \\ \hline
\end{longtable}
\subsection{UC15: Managing Platform Services}
\begin{longtable}{|>{\raggedright\arraybackslash}p{3.5cm}|>{\raggedright\arraybackslash}p{9.5cm}|}
\hline
\textbf{Attribute} & \textbf{Details} \\ \hline
\textbf{Actor} & Platform Manager \\ \hline
\textbf{Use Case} & Managing Platform Services \\ \hline
\textbf{Description} & Platform manager oversees and manages the services offered on the platform. \\ \hline
\textbf{Pre-conditions} & Services are active on the platform. \\ \hline
\textbf{Triggers} & Regular management or response to service-related issues. \\ \hline
\textbf{Main Scenario} & 
\begin{enumerate}
    \item Manager logs into the management portal.
    \item Manager reviews and updates service listings.
    \item Manager addresses any service-related issues or feedback.
\end{enumerate} \\ \hline
\textbf{Post-conditions} & Platform services are effectively managed and updated. \\ \hline
\textbf{Extensions} & 
\begin{enumerate}
    \item Unauthorized access attempt.
    \item Technical issues in the management portal.
\end{enumerate} \\ \hline
\textbf{Relationships} & Central to Platform Operations and Service Quality. \\ \hline
\end{longtable}
\subsection{UC16: Managing User Profiles and Accounts}
\begin{longtable}{|>{\raggedright\arraybackslash}p{3.5cm}|>{\raggedright\arraybackslash}p{9.5cm}|}
\hline
\textbf{Attribute} & \textbf{Details} \\ \hline
\textbf{Actor} & User \\ \hline
\textbf{Use Case} & Managing User Profiles and Accounts \\ \hline
\textbf{Description} & User manages their personal profile and account settings. \\ \hline
\textbf{Pre-conditions} & User is registered and has an account. \\ \hline
\textbf{Triggers} & User needs to update profile information or account settings. \\ \hline
\textbf{Main Scenario} & 
\begin{enumerate}
    \item User logs into their account.
    \item User accesses the profile settings.
    \item User updates their profile or account settings as needed.
    \item Changes are saved and applied.
\end{enumerate} \\ \hline
\textbf{Post-conditions} & User's profile and account settings are updated. \\ \hline
\textbf{Extensions} & 
\begin{enumerate}
    \item Invalid input in profile update.
    \item Technical issues saving changes.
\end{enumerate} \\ \hline
\textbf{Relationships} & Integral to User Account Management. \\ \hline
\end{longtable}
\subsection{UC17: Managing Transactions and Payments}
\begin{longtable}{|>{\raggedright\arraybackslash}p{3.5cm}|>{\raggedright\arraybackslash}p{9.5cm}|}
\hline
\textbf{Attribute} & \textbf{Details} \\ \hline
\textbf{Actor} & User \\ \hline
\textbf{Use Case} & Managing Transactions and Payments \\ \hline
\textbf{Description} & User manages their financial transactions and payment methods. \\ \hline
\textbf{Pre-conditions} & User has conducted or wishes to conduct transactions. \\ \hline
\textbf{Triggers} & User needs to review or manage transactions or payment methods. \\ \hline
\textbf{Main Scenario} & 
\begin{enumerate}
    \item User logs into their account.
    \item User accesses the transactions section.
    \item User reviews transaction history or manages payment methods.
    \item Any changes or updates are saved.
\end{enumerate} \\ \hline
\textbf{Post-conditions} & User's transactions and payment methods are managed. \\ \hline
\textbf{Extensions} & 
\begin{enumerate}
    \item Transaction dispute.
    \item Technical issues with payment processing.
\end{enumerate} \\ \hline
\textbf{Relationships} & Linked with Financial Management and Security. \\ \hline
\end{longtable}
\subsection{UC18: Connecting with Transport Companies}
\begin{longtable}{|>{\raggedright\arraybackslash}p{3.5cm}|>{\raggedright\arraybackslash}p{9.5cm}|}
\hline
\textbf{Attribute} & \textbf{Details} \\ \hline
\textbf{Actor} & User \\ \hline
\textbf{Use Case} & Connecting with Transport Companies \\ \hline
\textbf{Description} & User connects with transport companies for waste transport services. \\ \hline
\textbf{Pre-conditions} & User requires waste transport services. \\ \hline
\textbf{Triggers} & User decides to arrange for waste transport. \\ \hline
\textbf{Main Scenario} & 
\begin{enumerate}
    \item User accesses the list of transport companies.
    \item User selects a company based on their requirements.
    \item User contacts the company or books a service directly.
\end{enumerate} \\ \hline
\textbf{Post-conditions} & User is connected with a transport company for waste transport. \\ \hline
\textbf{Extensions} & 
\begin{enumerate}
    \item No suitable transport companies available.
    \item Technical issues during booking.
\end{enumerate} \\ \hline
\textbf{Relationships} & Supports Waste Management and Transportation Services. \\ \hline
\end{longtable}
\subsection{UC19: Viewing and Analyzing Waste Management Data}
\begin{longtable}{|>{\raggedright\arraybackslash}p{3.5cm}|>{\raggedright\arraybackslash}p{9.5cm}|}
\hline
\textbf{Attribute} & \textbf{Details} \\ \hline
\textbf{Actor} & Platform Manager \\ \hline
\textbf{Use Case} & Viewing and Analyzing Waste Management Data \\ \hline
\textbf{Description} & Platform manager views and analyzes data related to waste management. \\ \hline
\textbf{Pre-conditions} & Data is collected and stored on the platform. \\ \hline
\textbf{Triggers} & Need for data analysis or reporting. \\ \hline
\textbf{Main Scenario} & 
\begin{enumerate}
    \item Manager accesses the data analytics dashboard.
    \item Manager reviews various metrics and reports.
    \item Manager uses insights for decision making and improvement.
\end{enumerate} \\ \hline
\textbf{Post-conditions} & Manager has an updated understanding of waste management performance. \\ \hline
\textbf{Extensions} & 
\begin{enumerate}
    \item Inaccurate or incomplete data.
    \item Technical issues with analytics tools.
\end{enumerate} \\ \hline
\textbf{Relationships} & Integral to Data-Driven Decision Making and Platform Strategy. \\ \hline
\end{longtable}
\subsection{UC20: Monitoring Recycling and Waste Management Performance}
\begin{longtable}{|>{\raggedright\arraybackslash}p{3.5cm}|>{\raggedright\arraybackslash}p{9.5cm}|}
\hline
\textbf{Attribute} & \textbf{Details} \\ \hline
\textbf{Actor} & Platform Manager \\ \hline
\textbf{Use Case} & Monitoring Recycling and Waste Management Performance \\ \hline
\textbf{Description} & Manager monitors and evaluates the performance of recycling and waste management. \\ \hline
\textbf{Pre-conditions} & Recycling and waste management processes are in operation. \\ \hline
\textbf{Triggers} & Regular performance review or specific analysis request. \\ \hline
\textbf{Main Scenario} & 
\begin{enumerate}
    \item Manager accesses performance metrics.
    \item Manager analyzes recycling rates and waste management effectiveness.
    \item Manager generates reports for stakeholders.
\end{enumerate} \\ \hline
\textbf{Post-conditions} & Performance of recycling and waste management is assessed. \\ \hline
\textbf{Extensions} & N/A \\ \hline
\textbf{Relationships} & N/A \\ \hline
\end{longtable}
\subsection{UC21: Managing Shipping and Delivery Details}
\begin{longtable}{|>{\raggedright\arraybackslash}p{3.5cm}|>{\raggedright\arraybackslash}p{9.5cm}|}
\hline
\textbf{Attribute} & \textbf{Details} \\ \hline
\textbf{Actor} & User \\ \hline
\textbf{Use Case} & Managing Shipping and Delivery Details \\ \hline
\textbf{Description} & User manages and tracks shipping and delivery details for their transactions. \\ \hline
\textbf{Pre-conditions} & User has engaged in transactions requiring shipping. \\ \hline
\textbf{Triggers} & User needs to set up or track shipping for a transaction. \\ \hline
\textbf{Main Scenario} & 
\begin{enumerate}
    \item User accesses their transaction history.
    \item User sets or updates shipping details.
    \item User tracks the delivery status.
\end{enumerate} \\ \hline
\textbf{Post-conditions} & User's shipping and delivery details are managed. \\ \hline
\textbf{Extensions} & 
\begin{enumerate}
    \item Incorrect shipping information.
    \item Delays or issues with delivery.
\end{enumerate} \\ \hline
\textbf{Relationships} & Linked with Transaction Processing and Logistics Management. \\ \hline
\end{longtable}
\section{Original Class Diagram PlantUML Code}\label{secB}
\begin{verbatim}
@startuml
!define RECTANGLE class

' Main Classes and Subclasses
RECTANGLE User {
    +UserID: string
    +Name: string
    +Email: string
    +PhoneNumber: string
    +Address: string
    +UserType: string
    +RegistrationDate: dateTime
    +AverageUserRating: decimal
}
RECTANGLE IndividualUser {
}
RECTANGLE CorporateUser {
}
RECTANGLE Product {
    +ProductID: string
    +ProductName: string
    +Description: string
    +Category: string
    +Quantity: integer
    +UnitPrice: decimal
    +ListingDate: dateTime
    +SellerID: string
    +QualityCertification: string
    +WasteReductionPercentage: decimal
    +EnergySaved: decimal
}
RECTANGLE Plastic {
}
RECTANGLE Paper {
}
RECTANGLE Metal {
}
RECTANGLE Glass {
}
RECTANGLE Transaction {
    +TransactionID: string
    +BuyerID: string
    +SellerIDTransaction: string
    +ProductIDTransaction: string
    +QuantityTransaction: integer
    +TotalPrice: decimal
    +TransactionDate: dateTime
}
RECTANGLE Review {
    +ReviewID: string
    +ReviewerID: string
    +ProductIDReview: string
    +Rating: integer
    +Comment: string
    +ReviewDate: dateTime
}
RECTANGLE CollectionPoint {
    +PointID: string
    +LocationCP: string
    +TypeCP: string
    +OperatingHoursCP: string
}
RECTANGLE RecyclingPoint {
    +PointIDRec: string
    +LocationRP: string
    +TypesAccepted: string
    +OperatingHoursRP: string
}
RECTANGLE ServiceRequest {
    +RequestID: string
    +UserIDSR: string
    +ServiceType: string
    +Status: string
    +DateSR: dateTime
    +LocationSR: string
}
RECTANGLE TransportCompany {
    +CompanyID: string
    +CompanyName: string
    +ServicesOffered: string
    +ContactDetailsTC: string
}
RECTANGLE StateAdministration {
    +AdministrationID: string
    +AdminName: string
    +TypeSA: string
    +ContactDetailsSA: string
}
RECTANGLE PaymentGateway {
    +PaymentID: string
    +PaymentMethod: string
    +Amount: decimal
    +PaymentDate: dateTime
    +PaymentConfirmationCode: string
}
RECTANGLE EnvironmentalImpact {
    +ImpactID: string
    +ProductIDEnv: string
    +CarbonFootprint: decimal
    +RecyclingEfficiency: decimal
    +EcoLabel: string
}
RECTANGLE InformationResource {
    +ResourceID: string
    +Title: string
    +TypeIR: string
    +Content: string
    +DatePublished: dateTime
}

RECTANGLE RewardSystem {
    +RewardID: string
    +UserIDRew: string
    +PointsEarned: integer
    +RedeemableItems: string
    +DateEarned: dateTime
}
RECTANGLE PaymentDetails {
    +PaymentID: string
    +PaymentMethod: string
    +Amount: decimal
    +PaymentDate: dateTime
    +PaymentConfirmationCode: string
}
RECTANGLE ShippingDetails {
    +ShippingID: string
    +TransactionID: string
    +Address: string
    +ExpectedDeliveryDate: dateTime
    +ShippingStatus: string
    +Carrier: string
    +EstimatedShippingTime: duration
}

' Subclassing
IndividualUser --|> User
CorporateUser --|> User
Plastic --|> Product
Paper --|> Product
Metal --|> Product
Glass --|> Product

' Relationships and Associations
User "1" -- "0..*" Product : hasProduct
User "1" -- "0..*" Transaction : hasTransaction
User "1" -- "0..*" Review : hasReview
Product "1" -- "0..1" EnvironmentalImpact : hasEnvironmentalImpact
User "1" -- "0..*" CollectionPoint : hasCollectionPoint
TransportCompany "1" -- "0..*" ServiceRequest : providesServiceRequest
StateAdministration "1" -- "0..*" Transaction : consults
StateAdministration "1" -- "0..*" ServiceRequest : consults
PaymentGateway "1" -- "0..*" Transaction : managesPayment
User "1" -- "0..*" InformationResource : hasInformationResource
User "1" -- "0..*" RewardSystem : hasRewardSystem
Transaction "1" -- "0..1" PaymentDetails : hasPaymentDetails
Transaction "1" -- "0..1" ShippingDetails : hasShippingDetails


@enduml
\end{verbatim}

\section{Enhanced Class Diagram PlantUML Code}\label{secA2}
\begin{verbatim}
    
@startuml
!define RECTANGLE class

RECTANGLE User {
    +UserID: string
    +Name: string
    +Email: string
    +PhoneNumber: string
    +Address: string
    +UserType: string
    +RegistrationDate: dateTime
    +AverageUserRating: decimal
    +registerUser(name: string, email: string, phoneNumber: string, address: string): boolean
    +listProduct(productDetails: ProductDetails): boolean
    +manageRewards(): string
    +trackWasteJourney(trackingCode: string): string
    +submitFeedbackOrReport(feedbackDetails: FeedbackDetails): boolean
    +updateProfile(profileDetails: ProfileDetails): boolean
    +manageTransactionsAndPayments(): void
    +connectWithTransportCompanies(): string
    +manageShippingAndDeliveryDetails(transactionID: 
    string, shippingDetails: ShippingDetails): boolean
}
RECTANGLE IndividualUser {
}
RECTANGLE CorporateUser {
}
RECTANGLE Product {
    +ProductID: string
    +ProductName: string
    +Description: string
    +Category: string
    +Quantity: integer
    +UnitPrice: decimal
    +ListingDate: dateTime
    +SellerID: string
    +QualityCertification: string
    +WasteReductionPercentage: decimal
    +EnergySaved: decimal
}
RECTANGLE Plastic {
}
RECTANGLE Paper {
}
RECTANGLE Metal {
}
RECTANGLE Glass {
}
RECTANGLE Transaction {
    +TransactionID: string
    +BuyerID: string
    +SellerIDTransaction: string
    +ProductIDTransaction: string
    +QuantityTransaction: integer
    +TotalPrice: decimal
    +TransactionDate: dateTime
    +processTransaction(transactionDetails: TransactionDetails): boolean
    +processSale(saleDetails: SaleDetails): boolean
    +trackWasteJourney(trackingCode: string): string
    +manageShippingAndDeliveryDetails(transactionID: string,
    shippingDetails: ShippingDetails): boolean
}
RECTANGLE Review {
    +ReviewID: string
    +ReviewerID: string
    +ProductIDReview: string
    +Rating: integer
    +Comment: string
    +ReviewDate: dateTime
    +submitReview(reviewDetails: ReviewDetails): boolean
}
RECTANGLE CollectionPoint {
    +PointID: string
    +LocationCP: string
    +TypeCP: string
    +OperatingHoursCP: string
}
RECTANGLE RecyclingPoint {
    +PointIDRec: string
    +LocationRP: string
    +TypesAccepted: string
    +OperatingHoursRP: string
}
RECTANGLE ServiceRequest {
    +RequestID: string
    +UserIDSR: string
    +ServiceType: string
    +Status: string
    +DateSR: dateTime
    +LocationSR: string
    +submitCollectionRequest(requestDetails: CollectionRequestDetails): boolean
    +submitTransportRequest(requestDetails: TransportRequestDetails): boolean
    +getServiceRequestDetails(requestID: string): string
}
RECTANGLE TransportCompany {
    +CompanyID: string
    +CompanyName: string
    +ServicesOffered: string
    +ContactDetailsTC: string
}
RECTANGLE StateAdministration {
    +AdministrationID: string
    +AdminName: string
    +TypeSA: string
    +ContactDetailsSA: string
}
RECTANGLE PaymentGateway {
    +PaymentID: string
    +PaymentMethod: string
    +Amount: decimal
    +PaymentDate: dateTime
    +PaymentConfirmationCode: string
    +processTransaction(transactionDetails: TransactionDetails): boolean
    +manageTransactionsAndPayments(): void
}
RECTANGLE EnvironmentalImpact {
    +ImpactID: string
    +ProductIDEnv: string
    +CarbonFootprint: decimal
    +RecyclingEfficiency: decimal
    +EcoLabel: string
}
RECTANGLE InformationResource {
    +ResourceID: string
    +Title: string
    +TypeIR: string
    +Content: string
    +DatePublished: dateTime
}

RECTANGLE RewardSystem {
    +RewardID: string
    +UserIDRew: string
    +PointsEarned: integer
    +RedeemableItems: string
    +DateEarned: dateTime
}
RECTANGLE PaymentDetails {
    +PaymentID: string
    +PaymentMethod: string
    +Amount: decimal
    +PaymentDate: dateTime
    +PaymentConfirmationCode: string
}
RECTANGLE ShippingDetails {
    +ShippingID: string
    +TransactionID: string
    +Address: string
    +ExpectedDeliveryDate: dateTime
    +ShippingStatus: string
    +Carrier: string
    +EstimatedShippingTime: duration
}

RECTANGLE PlatformManager {
    +ManagerID: string
    +Name: string
    +Role: string
    +login(username: string, password: string): boolean
    +viewAndAnalyzeWasteManagementData(): string
    +monitorRecyclingAndWasteManagementPerformance(): string
}

' Subclassing
IndividualUser --|> User
CorporateUser --|> User
Plastic --|> Product
Paper --|> Product
Metal --|> Product
Glass --|> Product

' Relationships and Associations
User "1" -- "0..*" Product : hasProduct
User "1" -- "0..*" Transaction : hasTransaction
User "1" -- "0..*" Review : hasReview
Product "1" -- "0..1" EnvironmentalImpact : hasEnvironmentalImpact
User "1" -- "0..*" CollectionPoint : hasCollectionPoint
User "1" -- "0..*" RecyclingPoint : hasRecyclingPoint
TransportCompany "1" -- "0..*" ServiceRequest : providesServiceRequest
StateAdministration "1" -- "0..*" Transaction : consults
StateAdministration "1" -- "0..*" ServiceRequest : consults
PaymentGateway "1" -- "0..*" Transaction : managesPayment
PaymentGateway "1" -- "0..*" Transaction : processTransaction
User "1" -- "0..*" InformationResource : hasInformationResource
User "1" -- "0..*" RewardSystem : hasRewardSystem
Transaction "1" -- "0..1" PaymentDetails : hasPaymentDetails
Transaction "1" -- "0..1" ShippingDetails : hasShippingDetails

@enduml
\end{verbatim}

\end{appendices}

\bibliography{sn-bibliography}


\begin{thebibliography}{42}
\ifx \bisbn   \undefined \def \bisbn  #1{ISBN #1}\fi
\ifx \binits  \undefined \def \binits#1{#1}\fi
\ifx \bauthor  \undefined \def \bauthor#1{#1}\fi
\ifx \batitle  \undefined \def \batitle#1{#1}\fi
\ifx \bjtitle  \undefined \def \bjtitle#1{#1}\fi
\ifx \bvolume  \undefined \def \bvolume#1{\textbf{#1}}\fi
\ifx \byear  \undefined \def \byear#1{#1}\fi
\ifx \bissue  \undefined \def \bissue#1{#1}\fi
\ifx \bfpage  \undefined \def \bfpage#1{#1}\fi
\ifx \blpage  \undefined \def \blpage #1{#1}\fi
\ifx \burl  \undefined \def \burl#1{\textsf{#1}}\fi
\ifx \doiurl  \undefined \def \doiurl#1{\url{https://doi.org/#1}}\fi
\ifx \betal  \undefined \def \betal{\textit{et al.}}\fi
\ifx \binstitute  \undefined \def \binstitute#1{#1}\fi
\ifx \binstitutionaled  \undefined \def \binstitutionaled#1{#1}\fi
\ifx \bctitle  \undefined \def \bctitle#1{#1}\fi
\ifx \beditor  \undefined \def \beditor#1{#1}\fi
\ifx \bpublisher  \undefined \def \bpublisher#1{#1}\fi
\ifx \bbtitle  \undefined \def \bbtitle#1{#1}\fi
\ifx \bedition  \undefined \def \bedition#1{#1}\fi
\ifx \bseriesno  \undefined \def \bseriesno#1{#1}\fi
\ifx \blocation  \undefined \def \blocation#1{#1}\fi
\ifx \bsertitle  \undefined \def \bsertitle#1{#1}\fi
\ifx \bsnm \undefined \def \bsnm#1{#1}\fi
\ifx \bsuffix \undefined \def \bsuffix#1{#1}\fi
\ifx \bparticle \undefined \def \bparticle#1{#1}\fi
\ifx \barticle \undefined \def \barticle#1{#1}\fi
\bibcommenthead
\ifx \bconfdate \undefined \def \bconfdate #1{#1}\fi
\ifx \botherref \undefined \def \botherref #1{#1}\fi
\ifx \url \undefined \def \url#1{\textsf{#1}}\fi
\ifx \bchapter \undefined \def \bchapter#1{#1}\fi
\ifx \bbook \undefined \def \bbook#1{#1}\fi
\ifx \bcomment \undefined \def \bcomment#1{#1}\fi
\ifx \oauthor \undefined \def \oauthor#1{#1}\fi
\ifx \citeauthoryear \undefined \def \citeauthoryear#1{#1}\fi
\ifx \endbibitem  \undefined \def \endbibitem {}\fi
\ifx \bconflocation  \undefined \def \bconflocation#1{#1}\fi
\ifx \arxivurl  \undefined \def \arxivurl#1{\textsf{#1}}\fi
\csname PreBibitemsHook\endcsname

\bibitem[\protect\citeauthoryear{Team}{2021}]{plantuml}
\begin{botherref}
\oauthor{\bsnm{Team}, \binits{P.}}:
PlantUML.
\url{https://plantuml.com},
Paris, France
(2021)
\end{botherref}
\endbibitem

\bibitem[\protect\citeauthoryear{Ali}{2023}]{Ali2023}
\begin{barticle}
\bauthor{\bsnm{Ali}, \binits{J.M.}}:
\batitle{Ai-driven software engineering}.
\bjtitle{Advances in Engineering Innovation}
(\byear{2023})
\doiurl{10.54254/2977-3903/3/2023030}
\end{barticle}
\endbibitem

\bibitem[\protect\citeauthoryear{A. et~al.}{2021}]{batarseh2021}
\begin{barticle}
\bauthor{\bsnm{A.}, \binits{B.F.}},
\bauthor{\bsnm{Rasika}, \binits{M.}},
\bauthor{\bsnm{Abhinav}, \binits{K.}},
\bauthor{\bsnm{Justin}, \binits{B.}}:
\batitle{The application of artificial intelligence in software engineering: A review challenging conventional wisdom}.
\bjtitle{CoRR}
(\byear{2021})
\doiurl{10.48550/arXiv.2108.01591}
{\href{https://arxiv.org/abs/2108.01591}{{arXiv:2108.01591}}}
{[cs.SE]}
\end{barticle}
\endbibitem

\bibitem[\protect\citeauthoryear{Zhao et~al.}{2021}]{Zhao2021}
\begin{barticle}
\bauthor{\bsnm{Zhao}, \binits{L.}},
\bauthor{\bsnm{Alhoshan}, \binits{W.}},
\bauthor{\bsnm{Ferrari}, \binits{A.}},
\bauthor{\bsnm{Letsholo}, \binits{K.J.}},
\bauthor{\bsnm{Ajagbe}, \binits{M.}},
\bauthor{\bsnm{Chioasca}, \binits{E.-V.}},
\bauthor{\bsnm{Batista-Navarro}, \binits{R.T.}}:
\batitle{Natural language processing for requirements engineering: A systematic mapping study}.
\bjtitle{ACM Computing Surveys}
\bvolume{54}(\bissue{3}),
\bfpage{1}--\blpage{41}
(\byear{2021})
\doiurl{10.1145/3444689}
\end{barticle}
\endbibitem

\bibitem[\protect\citeauthoryear{Wang}{2023}]{Wang2023}
\begin{botherref}
\oauthor{\bsnm{Wang}, \binits{L.}}:
Ai in software engineering: Case studies and prospects
(2023)
\doiurl{10.48550/arXiv.2309.15768}
{\href{https://arxiv.org/abs/2309.15768}{{arXiv:2309.15768}}}
{[cs.SE]}
\end{botherref}
\endbibitem

\bibitem[\protect\citeauthoryear{Martinez-Fernandez et~al.}{2022}]{Martinez-Fernandez2022}
\begin{barticle}
\bauthor{\bsnm{Martinez-Fernandez}, \binits{S.}},
\bauthor{\bsnm{Bogner}, \binits{J.}},
\bauthor{\bsnm{Franch}, \binits{X.}},
\bauthor{\bsnm{Oriol}, \binits{M.}},
\bauthor{\bsnm{Siebert}, \binits{J.}},
\bauthor{\bsnm{Trendowicz}, \binits{A.}},
\bauthor{\bsnm{Vollmer}, \binits{A.M.}},
\bauthor{\bsnm{Wagner}, \binits{S.}}:
\batitle{Software engineering for ai-based systems: A survey}.
\bjtitle{ACM Transactions on Software Engineering and Methodology}
\bvolume{31}(\bissue{2}),
\bfpage{1}--\blpage{59}
(\byear{2022})
\doiurl{10.1145/3487043}
\end{barticle}
\endbibitem

\bibitem[\protect\citeauthoryear{Khurana et~al.}{2023}]{Khurana2023}
\begin{barticle}
\bauthor{\bsnm{Khurana}, \binits{D.}},
\bauthor{\bsnm{Koli}, \binits{A.}},
\bauthor{\bsnm{Khatter}, \binits{K.}},
\bauthor{\bsnm{Singh}, \binits{S.}}:
\batitle{Natural language processing: State of the art, current trends and challenges}.
\bjtitle{Multimedia Tools and Applications}
\bvolume{82},
\bfpage{3713}--\blpage{3744}
(\byear{2023})
\doiurl{10.1007/s11042-022-13428-4}
\end{barticle}
\endbibitem

\bibitem[\protect\citeauthoryear{Berardi et~al.}{2005}]{Berardi2005}
\begin{barticle}
\bauthor{\bsnm{Berardi}, \binits{D.}},
\bauthor{\bsnm{Calvanese}, \binits{D.}},
\bauthor{\bsnm{Giacomo}, \binits{G.D.}}:
\batitle{Reasoning on uml class diagrams}.
\bjtitle{Artif. Intell.}
\bvolume{168},
\bfpage{70}--\blpage{118}
(\byear{2005})
\doiurl{10.1016/j.artint.2005.05.003}
\end{barticle}
\endbibitem

\bibitem[\protect\citeauthoryear{Gray and Rumpe}{2020}]{Gray2020}
\begin{barticle}
\bauthor{\bsnm{Gray}, \binits{J.}},
\bauthor{\bsnm{Rumpe}, \binits{B.}}:
\batitle{Modeling dynamic structures}.
\bjtitle{Software \& Systems Modeling}
\bvolume{19},
\bfpage{527}--\blpage{528}
(\byear{2020})
\doiurl{10.1007/s10270-020-00793-7}
\end{barticle}
\endbibitem

\bibitem[\protect\citeauthoryear{Sajjii et~al.}{2023}]{Sajjii2023}
\begin{barticle}
\bauthor{\bsnm{Sajjii}, \binits{A.}},
\bauthor{\bsnm{Rhazali}, \binits{Y.}},
\bauthor{\bsnm{Hadi}, \binits{Y.}}:
\batitle{A methodology of automatic class diagrams generation from source code using model-driven architecture and machine learning to achieve energy efficiency}.
\bjtitle{E3S Web of Conferences}
\bvolume{412},
\bfpage{01002}
(\byear{2023})
\doiurl{10.1051/e3sconf/202341201002}
\end{barticle}
\endbibitem

\bibitem[\protect\citeauthoryear{Ahmed et~al.}{2022}]{Ahmed2022}
\begin{bchapter}
\bauthor{\bsnm{Ahmed}, \binits{S.}},
\bauthor{\bsnm{Ahmed}, \binits{A.}},
\bauthor{\bsnm{Eisty}, \binits{N.U.}}:
\bctitle{Automatic transformation of natural to unified modeling language: A systematic review}.
In: \bbtitle{2022 IEEE/ACIS 20th International Conference on Software Engineering Research, Management and Applications (SERA)}.
\bpublisher{IEEE}, \blocation{???}
(\byear{2022}).
\doiurl{10.1109/sera54885.2022.9806783} .
\burl{http://dx.doi.org/10.1109/SERA54885.2022.9806783}
\end{bchapter}
\endbibitem

\bibitem[\protect\citeauthoryear{Anda and Sjøberg}{2005}]{Anda2005}
\begin{barticle}
\bauthor{\bsnm{Anda}, \binits{B.}},
\bauthor{\bsnm{Sjøberg}, \binits{D.I.K.}}:
\batitle{Investigating the role of use cases in the construction of class diagrams}.
\bjtitle{Empirical Software Engineering}
\bvolume{10}(\bissue{3}),
\bfpage{285}--\blpage{309}
(\byear{2005})
\doiurl{10.1007/s10664-005-1289-3}
\end{barticle}
\endbibitem

\bibitem[\protect\citeauthoryear{OpenAI}{2021}]{OpenAI2021}
\begin{botherref}
\oauthor{\bsnm{OpenAI}}:
ChatGPT: A large-scale generative model for open-domain chat
(2021).
\url{https://github.com/openai/gpt-3}
\end{botherref}
\endbibitem

\bibitem[\protect\citeauthoryear{Perez et~al.}{2021}]{Perez2021}
\begin{botherref}
\oauthor{\bsnm{Perez}, \binits{L.}},
\oauthor{\bsnm{Ottens}, \binits{L.}},
\oauthor{\bsnm{Viswanathan}, \binits{S.}}:
Automatic Code Generation using Pre-Trained Language Models
(2021).
\doiurl{10.48550/arXiv.2102.10535} .
\url{https://arxiv.org/abs/2102.10535}
\end{botherref}
\endbibitem

\bibitem[\protect\citeauthoryear{Elena}{2014}]{Elena2014}
\begin{bchapter}
\bauthor{\bsnm{Elena}, \binits{C.}}:
\bctitle{An approach to class diagram design}.
In: \bbtitle{2014 2nd International Conference on Model-Driven Engineering and Software Development (MODELSWARD)},
pp. \bfpage{448}--\blpage{453}
(\byear{2014}).
\doiurl{10.5220/0004763504480453} .
\bcomment{IEEE}
\end{bchapter}
\endbibitem

\bibitem[\protect\citeauthoryear{Egyed}{2002}]{Egyed2002}
\begin{bchapter}
\bauthor{\bsnm{Egyed}, \binits{A.}}:
\bctitle{Automated abstraction of class diagrams},
vol. \bseriesno{11},
pp. \bfpage{449}--\blpage{491}
(\byear{2002}).
\doiurl{10.1145/606612.606616}
\end{bchapter}
\endbibitem

\bibitem[\protect\citeauthoryear{Guéhéneuc}{2004}]{Gueheneuc2004}
\begin{bchapter}
\bauthor{\bsnm{Guéhéneuc}, \binits{Y.-G.}}:
\bctitle{A reverse engineering tool for precise class diagrams},
pp. \bfpage{28}--\blpage{41}
(\byear{2004}).
\doiurl{10.1145/1034914.1034917}
\end{bchapter}
\endbibitem

\bibitem[\protect\citeauthoryear{Huang et~al.}{2016}]{Huang2016}
\begin{bchapter}
\bauthor{\bsnm{Huang}, \binits{L.}},
\bauthor{\bsnm{Duan}, \binits{Y.}},
\bauthor{\bsnm{Sun}, \binits{X.}},
\bauthor{\bsnm{Lin}, \binits{Z.}},
\bauthor{\bsnm{Zhu}, \binits{C.}}:
\bctitle{Enhancing uml class diagram abstraction with knowledge graph},
pp. \bfpage{606}--\blpage{616}
(\byear{2016}).
\doiurl{10.1007/978-3-319-46257-8_65}
\end{bchapter}
\endbibitem

\bibitem[\protect\citeauthoryear{Eichelberger}{2002}]{Eichelberger2002}
\begin{bchapter}
\bauthor{\bsnm{Eichelberger}, \binits{H.}}:
\bctitle{Aesthetics and automatic layout of uml class diagrams}.
In: \bbtitle{Proceedings First International Workshop on Visualizing Software for Understanding and Analysis},
pp. \bfpage{109}--\blpage{118}
(\byear{2002}).
\doiurl{10.1109/VISSOF.2002.1019791} .
\bcomment{ACM}
\end{bchapter}
\endbibitem

\bibitem[\protect\citeauthoryear{Sharma et~al.}{2010}]{Sharma2010}
\begin{bchapter}
\bauthor{\bsnm{Sharma}, \binits{V.S.}},
\bauthor{\bsnm{Sarkar}, \binits{S.}},
\bauthor{\bsnm{Verma}, \binits{K.}},
\bauthor{\bsnm{Panayappan}, \binits{A.}},
\bauthor{\bsnm{Kass}, \binits{A.}}:
\bctitle{Extracting high-level functional design from software requirements},
pp. \bfpage{35}--\blpage{42}
(\byear{2010}).
\doiurl{10.1109/APSEC.2009.63}
\end{bchapter}
\endbibitem

\bibitem[\protect\citeauthoryear{Osman et~al.}{2014}]{Osman2014}
\begin{barticle}
\bauthor{\bsnm{Osman}, \binits{H.}},
\bauthor{\bsnm{Hammad}, \binits{A.}},
\bauthor{\bsnm{El-Said}, \binits{M.}}:
\batitle{Using machine learning to condense class diagrams}.
\bjtitle{Journal of Systems and Software}
\bvolume{89},
\bfpage{113}--\blpage{126}
(\byear{2014})
\doiurl{10.1109/WICT.2014.7077321}
\end{barticle}
\endbibitem

\bibitem[\protect\citeauthoryear{Alashqar}{2021}]{Alashqar2021}
\begin{barticle}
\bauthor{\bsnm{Alashqar}, \binits{A.M.}}:
\batitle{Automatic generation of uml diagrams from scenario-based user requirements}.
\bjtitle{Jordanian Journal of Computers and Information Technology (JJCIT)}
\bvolume{07}(\bissue{02}),
\bfpage{180}--\blpage{191}
(\byear{2021})
\doiurl{10.5455/jjcit.71-1616087318}
\end{barticle}
\endbibitem

\bibitem[\protect\citeauthoryear{Nasiri et~al.}{2020}]{Nasiri2020}
\begin{barticle}
\bauthor{\bsnm{Nasiri}, \binits{S.}},
\bauthor{\bsnm{Rhazali}, \binits{Y.}},
\bauthor{\bsnm{Lahmer}, \binits{M.}},
\bauthor{\bsnm{Chenfour}, \binits{N.}}:
\batitle{Towards a generation of class diagram from user stories in agile methods}.
\bjtitle{Procedia Computer Science}
\bvolume{170},
\bfpage{831}--\blpage{837}
(\byear{2020})
\doiurl{10.1016/j.procs.2020.03.148} .
\bcomment{The 11th International Conference on Ambient Systems, Networks and Technologies (ANT) / The 3rd International Conference on Emerging Data and Industry 4.0 (EDI40) / Affiliated Workshops}
\end{barticle}
\endbibitem

\bibitem[\protect\citeauthoryear{Kumar and Sanyal}{2008}]{Kumar2008}
\begin{bchapter}
\bauthor{\bsnm{Kumar}, \binits{D.D.}},
\bauthor{\bsnm{Sanyal}, \binits{R.}}:
\bctitle{Static uml model generator from analysis of requirements (sugar)}.
In: \bbtitle{2008 Advanced Software Engineering and Its Applications},
pp. \bfpage{77}--\blpage{84}
(\byear{2008}).
\doiurl{10.1109/ASEA.2008.25}
\end{bchapter}
\endbibitem

\bibitem[\protect\citeauthoryear{Lucassen et~al.}{2017}]{Lucassen2017}
\begin{barticle}
\bauthor{\bsnm{Lucassen}, \binits{G.}},
\bauthor{\bsnm{Robeer}, \binits{M.}},
\bauthor{\bsnm{Dalpiaz}, \binits{F.}},
\bauthor{\bsnm{Werf}, \binits{J.M.E.M.}},
\bauthor{\bsnm{Brinkkemper}, \binits{S.}}:
\batitle{Extracting conceptual models from user stories with visual narrator}.
\bjtitle{Requirements Engineering}
\bvolume{22}(\bissue{3}),
\bfpage{339}--\blpage{358}
(\byear{2017})
\doiurl{10.1007/s00766-017-0270-1}
\end{barticle}
\endbibitem

\bibitem[\protect\citeauthoryear{Tantithamthavorn et~al.}{2020}]{Tantithamthavorn2020}
\begin{botherref}
\oauthor{\bsnm{Tantithamthavorn}, \binits{C.}},
\oauthor{\bsnm{Jiarpakdee}, \binits{J.}},
\oauthor{\bsnm{Grundy}, \binits{J.}}:
Explainable AI for Software Engineering
(2020).
\doiurl{10.48550/arXiv.2012.01614}
\end{botherref}
\endbibitem

\bibitem[\protect\citeauthoryear{Sergievskiy and Kirpichnikova}{2018}]{Sergievskiy2018}
\begin{barticle}
\bauthor{\bsnm{Sergievskiy}, \binits{M.}},
\bauthor{\bsnm{Kirpichnikova}, \binits{K.}}:
\batitle{Optimizing uml class diagrams}.
\bjtitle{ITM Web of Conferences}
\bvolume{18},
\bfpage{03003}
(\byear{2018})
\doiurl{10.1051/itmconf/20181803003}
\end{barticle}
\endbibitem

\bibitem[\protect\citeauthoryear{Eiglsperger and Siebenhaller}{2003}]{Eiglsperger2004}
\begin{barticle}
\bauthor{\bsnm{Eiglsperger}, \binits{M.}},
\bauthor{\bsnm{Siebenhaller}, \binits{M.}}:
\batitle{A topology-shape-metrics approach for the automatic layout of uml class diagrams}.
\bjtitle{Proceedings of ACM Symposium on Software Visualization}
(\byear{2003})
\doiurl{10.1145/774833.774860}
\end{barticle}
\endbibitem

\bibitem[\protect\citeauthoryear{Ha and Kim}{2013}]{Ha2013}
\begin{barticle}
\bauthor{\bsnm{Ha}, \binits{I.}},
\bauthor{\bsnm{Kim}, \binits{C.}}:
\batitle{Enhancement of consistency of uml diagrams by cross checking rules}.
\bjtitle{Applied Mechanics and Materials}
\bvolume{284-287},
\bfpage{3409}--\blpage{3412}
(\byear{2013})
\doiurl{10.4028/www.scientific.net/AMM.284-287.3409}
\end{barticle}
\endbibitem

\bibitem[\protect\citeauthoryear{Briand et~al.}{2012}]{Briand2012}
\begin{bchapter}
\bauthor{\bsnm{Briand}, \binits{L.}},
\bauthor{\bsnm{Labiche}, \binits{Y.}},
\bauthor{\bsnm{Liu}, \binits{Y.}}:
\bctitle{Combining uml sequence and state machine diagrams for data-flow based integration testing}.
In: \beditor{\bsnm{Vallecillo}, \binits{A.}},
\beditor{\bsnm{Tolvanen}, \binits{J.-P.}},
\beditor{\bsnm{Kindler}, \binits{E.}},
\beditor{\bsnm{St{\"o}rrle}, \binits{H.}},
\beditor{\bsnm{Kolovos}, \binits{D.}} (eds.)
\bbtitle{Modelling Foundations and Applications},
pp. \bfpage{74}--\blpage{89}.
\bpublisher{Springer},
\blocation{Berlin, Heidelberg}
(\byear{2012}).
\doiurl{10.1007/978-3-642-31491-9_8} .
\burl{https://doi.org/10.1007/978-3-642-31491-9\_8}
\end{bchapter}
\endbibitem

\bibitem[\protect\citeauthoryear{Aoumeur and Saake}{2002}]{Aoumeur2002}
\begin{bchapter}
\bauthor{\bsnm{Aoumeur}, \binits{N.}},
\bauthor{\bsnm{Saake}, \binits{G.}}:
\bctitle{Integrating and rapid-prototyping uml structural and behavioural diagrams using rewriting logic},
vol. \bseriesno{2348},
pp. \bfpage{296}--\blpage{310}
(\byear{2002}).
\doiurl{10.1007/3-540-47961-9_22}
\end{bchapter}
\endbibitem

\bibitem[\protect\citeauthoryear{Abdelnabi et~al.}{2020}]{Abdelnabi2020}
\begin{bchapter}
\bauthor{\bsnm{Abdelnabi}, \binits{E.A.}},
\bauthor{\bsnm{Maatuk}, \binits{A.M.}},
\bauthor{\bsnm{Abdelaziz}, \binits{T.M.}},
\bauthor{\bsnm{Elakeili}, \binits{S.M.}}:
\bctitle{Generating uml class diagram using nlp techniques and heuristic rules}.
In: \bbtitle{2020 20th International Conference on Sciences and Techniques of Automatic Control and Computer Engineering (STA)},
pp. \bfpage{277}--\blpage{282}
(\byear{2020}).
\doiurl{10.1109/STA50679.2020.9329301}
\end{bchapter}
\endbibitem

\bibitem[\protect\citeauthoryear{M.~Maatuk and A.~Abdelnabi}{2021}]{Maatuk2021}
\begin{bchapter}
\bauthor{\bsnm{M.~Maatuk}, \binits{A.}},
\bauthor{\bsnm{A.~Abdelnabi}, \binits{E.}}:
\bctitle{Generating uml use case and activity diagrams using nlp techniques and heuristics rules}.
In: \bbtitle{International Conference on Data Science, E-Learning and Information Systems 2021}.
\bsertitle{DATA'21},
pp. \bfpage{271}--\blpage{277}.
\bpublisher{Association for Computing Machinery},
\blocation{New York, NY, USA}
(\byear{2021}).
\doiurl{10.1145/3460620.3460768} .
\burl{https://doi.org/10.1145/3460620.3460768}
\end{bchapter}
\endbibitem

\bibitem[\protect\citeauthoryear{Danėnas et~al.}{2020}]{P2020}
\begin{barticle}
\bauthor{\bsnm{Danėnas}, \binits{P.}},
\bauthor{\bsnm{Skersys}, \binits{T.}},
\bauthor{\bsnm{Butleris}, \binits{R.}}:
\batitle{Natural language processing-enhanced extraction of sbvr business vocabularies and business rules from uml use case diagrams}.
\bjtitle{Data \& Knowledge Engineering}
\bvolume{128},
\bfpage{101822}
(\byear{2020})
\doiurl{10.1016/j.datak.2020.101822}
\end{barticle}
\endbibitem

\bibitem[\protect\citeauthoryear{Dawood~Omer et~al.}{2021}]{Omer2021}
\begin{barticle}
\bauthor{\bsnm{Dawood~Omer}, \binits{O.S.}},
\bauthor{\bsnm{Sahraoui}, \binits{A.-E.-K.}},
\bauthor{\bsnm{Mahmoud}, \binits{M.M.E.}},
\bauthor{\bsnm{Babiker}, \binits{A.-E.-A.}}:
\batitle{Requirements and design consistency: A bi-directional traceability and natural language processing assisted approach}.
\bjtitle{European Journal of Engineering and Technology Research}
\bvolume{6}(\bissue{3}),
\bfpage{120}--\blpage{129}
(\byear{2021})
\doiurl{10.24018/ejeng.2021.6.3.2373}
\end{barticle}
\endbibitem

\bibitem[\protect\citeauthoryear{Dawood and Sahraoui}{2018}]{Dawood2018}
\begin{barticle}
\bauthor{\bsnm{Dawood}, \binits{O.S.}},
\bauthor{\bsnm{Sahraoui}, \binits{A.-E.-K.}}:
\batitle{Toward requirements and design traceability using natural language processing}.
\bjtitle{European Journal of Engineering and Technology Research}
\bvolume{3}(\bissue{7}),
\bfpage{42}--\blpage{49}
(\byear{2018})
\doiurl{10.24018/ejeng.2018.3.7.807}
\end{barticle}
\endbibitem

\bibitem[\protect\citeauthoryear{Arellano et~al.}{2015}]{Arellano2015}
\begin{barticle}
\bauthor{\bsnm{Arellano}, \binits{A.}},
\bauthor{\bsnm{Zontek-Carney}, \binits{E.}},
\bauthor{\bsnm{Austin}, \binits{M.}}:
\batitle{Frameworks for natural language processing of textual requirements}.
\bjtitle{International Journal on Advances in Systems and Measurements}
\bvolume{8},
\bfpage{230}--\blpage{240}
(\byear{2015})
\doiurl{10.1145/3384613.3384619}
\end{barticle}
\endbibitem

\bibitem[\protect\citeauthoryear{Yue et~al.}{2015}]{Yue2015}
\begin{botherref}
\oauthor{\bsnm{Yue}, \binits{T.}},
\oauthor{\bsnm{Briand}, \binits{L.C.}},
\oauthor{\bsnm{Labiche}, \binits{Y.}}:
atoucan: An automated framework to derive uml analysis models from use case models
\textbf{24}(3)
(2015)
\doiurl{10.1145/2699697}
\end{botherref}
\endbibitem

\bibitem[\protect\citeauthoryear{Salehi and Burgueño}{2018}]{Salehi2018}
\begin{barticle}
\bauthor{\bsnm{Salehi}, \binits{H.}},
\bauthor{\bsnm{Burgueño}, \binits{R.}}:
\batitle{Emerging artificial intelligence methods in structural engineering}.
\bjtitle{Engineering Structures}
\bvolume{171},
\bfpage{170}--\blpage{189}
(\byear{2018})
\doiurl{10.1016/j.engstruct.2018.05.084}
\end{barticle}
\endbibitem

\bibitem[\protect\citeauthoryear{Abdelnabi et~al.}{2021}]{Abdelnabi2021}
\begin{bchapter}
\bauthor{\bsnm{Abdelnabi}, \binits{E.}},
\bauthor{\bsnm{Maatuk}, \binits{A.}},
\bauthor{\bsnm{Hagal}, \binits{M.}}:
\bctitle{Generating uml class diagram from natural language requirements: A survey of approaches and techniques},
pp. \bfpage{288}--\blpage{293}
(\byear{2021}).
\doiurl{10.1109/MI-STA52233.2021.9464433}
\end{bchapter}
\endbibitem

\bibitem[\protect\citeauthoryear{Schwaber and Beedle}{2002}]{schwaber2002}
\begin{bbook}
\bauthor{\bsnm{Schwaber}, \binits{K.}},
\bauthor{\bsnm{Beedle}, \binits{M.}}:
\bbtitle{Agile Software Development with Scrum}.
\bpublisher{Prentice Hall}, \blocation{???}
(\byear{2002})
\end{bbook}
\endbibitem

\bibitem[\protect\citeauthoryear{Kniberg and Skarin}{2010}]{kniberg2010}
\begin{bbook}
\bauthor{\bsnm{Kniberg}, \binits{H.}},
\bauthor{\bsnm{Skarin}, \binits{M.}}:
\bbtitle{Kanban and Scrum: Making the Most of Both}.
\bpublisher{C4Media}, \blocation{???}
(\byear{2010})
\end{bbook}
\endbibitem

\end{thebibliography}

\end{document}